\documentclass[prd,aps,nofootinbib,preprint,tightenlines]{revtex4}

\usepackage{graphicx}
\usepackage{amsmath}
\usepackage{amsfonts}
\usepackage{amssymb}
\usepackage{bm} 
\usepackage[usenames]{color}

\newcommand{\beq}{\begin{equation}}
\newcommand{\eeq}{\end{equation}}
\newcommand{\bea}{\begin{eqnarray}}
\newcommand{\eea}{\end{eqnarray}}
\def\simge{\mathrel{
  \rlap{\raise 0.511ex \hbox{$>$}}{\lower 0.511ex \hbox{$\sim$}}}}
\def\simle{\mathrel{
  \rlap{\raise 0.511ex \hbox{$<$}}{\lower 0.511ex \hbox{$\sim$}}}}

\begin{document}

\title{Quarkonium states in a complex-valued potential}

\author{Matthew Margotta, Kyle McCarty, Christina McGahan, Michael Strickland, and David Yager-Elorriaga}
\affiliation{Department of Physics, Gettysburg College, Gettysburg, PA 17325, USA\\}

\begin{abstract}
We calculate quarkonium binding energies using a realistic complex-valued potential
for both an isotropic and anisotropic quark-gluon plasma.  We determine the disassociation
temperatures of the ground and first excited states considering both the real and
imaginary parts of the binding energy.  We show that the effect of momentum-space
anisotropy is smaller on the imaginary part of the binding energy than on the real
part of the binding energy.  In the case that one assumes an isotropic plasma,
we find disassociation temperatures for the $J/\psi$, $\Upsilon$ and $\chi_b$ of
1.6 $T_c$, 2.8 $T_c$, and 1.5 $T_c$, respectively.   We find that a finite oblate
momentum-space anisotropy increases the disassociation temperature for all 
states considered and results in a splitting of the p-wave states associated
with the $\chi_b$ first excited state of bottomonium.

\end{abstract}


\maketitle

\section{Introduction}

The behavior of nuclear matter at extreme temperatures is now being studied with the 
highest collision energies ever achieved using the Large Hadron Collider (LHC) at CERN.
The ultrarelativistic heavy ion collisions being studied there will eventually 
have a center of mass energy of 5.5 TeV per nucleon, which is 27.5 times higher than the
200 GeV per nucleon energy achieved at the Relativistic Heavy Ion Collider 
(RHIC) at Brookhaven National Laboratory.  At
RHIC, observations indicated that initial temperatures on the order of twice the critical
temperature for the quark-gluon plasma phase transition were generated.  This corresponds
to $T_0 \sim 360$ MeV.  Assuming that the initial temperature scales with the
fourth root of the collision energy as predicted by dimensional analysis, one predicts
that initial temperatures on the order of $T_0 \sim 4.6\,T_c \sim 830$ MeV
will be generated at the LHC.  At such high temperatures, one expects to generate
a quark-gluon plasma in which the formation of quark bound
states is suppressed in favor of a state of matter consisting of a deconfined plasma of quarks
and gluons.

Suppression of quark bound states follows from the fact that in the 
quark-gluon plasma one expects color charge to be Debye screened \cite{Shuryak:1980tp,Gross:1980br}.  
This effect led to early proposals to use heavy quarkonium production to measure the temperature
of the quark-gluon plasma.  Heavy quarkonium has received the most theoretical
attention since heavy quark states are dominated by short rather than
long distance physics at low temperatures and can be treated using heavy quark
effective theory.  Based on such effective theories of Quantum Chromodynamics (QCD) with weak coupling
at short distances, non-relativistic quarkonium states can be reliably described. Their binding
energies are much smaller than the quark mass $m_Q\gg\Lambda_{\rm
  QCD}$ ($Q=c,b$), and their sizes are much larger than $1/m_Q$. At zero
temperature, since the velocity of the quarks in the bound state is
small, $v\ll c$, quarkonium can be understood in terms of
non-relativistic potential models \cite{Lucha:1991vn} such as the
Cornell potential \cite{Eichten:1979ms}. Such potential models can
be derived directly from QCD as an effective field theory
(potential non-relativistic QCD - pNRQCD) by integrating out modes
above the scales $m_Q$ and then $m_Q v$, respectively
\cite{Brambilla:2004jw}.

As mentioned above, at high temperature the deconfined phase of QCD exhibits
screening of static color-electric fields. It is expected that this
screening leads to the dissociation of quarkonium states, which can
serve as a signal for the formation of a deconfined quark-gluon plasma
in heavy ion collisions~\cite{Matsui:1986dk}. Inspired by the success at
zero temperature, potential model descriptions have also been applied to
understand quarkonium properties at finite temperature. The pioneering
paper of Matsui and Satz \cite{Matsui:1986dk} was followed by the work of
Karsch, Mehr, and Satz \cite{Karsch:1987pv}, which presented the first
quantitative calculation of quarkonium properties at high temperature. 
In recent work, more involved calculations
of quarkonium spectral functions and meson current correlators
obtained from potential models have been performed
\cite{Mocsy:2004bv,Wong:2004zr,Mocsy:2005qw,Cabrera:2006wh,%
Mocsy:2007jz,Alberico:2007rg,Mocsy:2007yj,Mocsy:2008eg}. 
The results have been compared to first-principle QCD calculations
performed numerically on lattices \cite{Umeda:2002vr,Asakawa:2003re,%
Datta:2003ww,Aarts:2006nr,Hatsuda:2006zz,Jakovac:2006sf,Umeda:2007hy,Aarts:2007pk,%
Aarts:2010ek} which rely on the maximum entropy method 
\cite{Nakahara:1999vy,Asakawa:2000tr,Asakawa:2002xj}.

A summary and review of the current understanding of potential
models is presented in \cite{Mocsy:2008eg}, and different aspects of quarkonium
in collider experiments can be found in \cite{Abreu:2007kv,Rapp:2008tf}.
In recent years, the imaginary part of the potential due to Landau
damping has been calculated \cite{Laine:2006ns,Laine:2007gj,Beraudo:2007ky}.
Also, the derivation of potential models from QCD via effective field theory methods has
been extended to finite temperature~\cite{Brambilla:2008cx}. All of the aforementioned 
calculations, however, were performed with the assumption of an
isotropic thermal medium.

In the last few years there has been an interest in the effect of plasma momentum-space
anisotropies on quarkonium binding energies for both ground and excited
states \cite{Dumitru:2007hy,Dumitru:2009ni,Burnier:2009yu,Dumitru:2009fy,%
Noronha:2009ia,Philipsen:2009wg}.  The interest stems from the fact that
at early times a viscous quark-gluon plasma can have large momentum-space anisotropies
\cite{Israel:1976tn,Israel:1979wp,Baym:1984np,%
Muronga:2001zk,Muronga:2003ta,Martinez:2009mf,Florkowski:2010cf,Martinez:2010sc,Martinez:2010sd}.
Depending on the magnitude of the shear viscosity, these momentum-space
anisotropies can persist for a long time ($\sim$ 1 - 10 fm/c).  The first paper to consider the quarkonium
potential in a momentum-space anisotropic plasma \cite{Dumitru:2007hy} considered
only the real part of the potential; however, two recent works have extended the calculation to
include the imaginary part of the potential \cite{Burnier:2009yu,Dumitru:2009fy}.
In this paper we use the imaginary part of the potential derived in \cite{Dumitru:2009fy}
and include it in a phenomenological model of the heavy quarkonium potential.  
We then numerically solve the three-dimensional Schr\"odinger equation to find
the real and imaginary parts of the binding energies and full quantum wavefunctions
of the charmonium and bottomonium ground states, as well as the first excited state of
bottomonium.
We present data as a function of the temperature and are able to identify the
full effect of the isotropic and anisotropic potentials on these states.

We compare our results with a recent analytic estimate of the imaginary part of the binding energy by
Dumitru \cite{Dumitru:2010id}.
We show that, for an isotropic plasma, 
the imaginary part of the binding energy is approximately linear in the temperature for temperatures near the 
phase transition in agreement with Ref.~\cite{Dumitru:2010id}.  
However, in the case of the $J/\psi$, we find a significantly smaller slope for the 
imaginary part of the binding energy as a function of temperature than predicted
by Ref.~\cite{Dumitru:2010id}.  The discrepancy most likely arises from the fact that
Ref.~\cite{Dumitru:2010id} assumed Coulombic wavefunctions.  
The potential used here includes modifications at both intermediate and long ranges, which causes 
the numerical wavefunctions to not be well approximated by Coulombic wavefunctions.  In addition, our
wavefunctions are complex with the imaginary part
growing in magnitude as the temperature is increased.  This effect was ignored by
the assumption of Coulombic wavefunctions in Ref.~\cite{Dumitru:2010id}.  
We find that when the states
are small and dominated by the screened Coulomb potential, the imaginary
part of the binding energy increases approximately linearly with the temperature; however, as the size of the 
bound state increases, the scale set by the string tension dominates and the imaginary part 
of the binding energy increases more slowly with increasing temperature.

The structure of this paper is as follows:  In Sec.~\ref{sec:potmodel}, we review
the potential introduced in Ref.~\cite{Dumitru:2009ni} and extend it to include
the imaginary part of the potential derived in Ref.~\cite{Dumitru:2009fy}.  
In Sec.~\ref{sec:nummethod}, we
review the numerical method that we use to solve the three-dimensional 
Schr\"odinger equation.  In Sec.~\ref{sec:results}, we present our numerical 
results for the real and imaginary parts of the binding energies of the charmonium
and bottomonium ground states and first excited state of bottomonium.  In Sec.~\ref{sec:conc}, we
state our conclusions and give an outlook for future work.  Finally, in an appendix
we present numerical benchmarks and tests of the code used here in order to
demonstrate its convergence and applicability to the problem at hand.

\section{Setup and Model Potential}
\label{sec:potmodel}

In this section we specify the potential we use in this work.
We consider the general case of a quark-gluon plasma which is anisotropic in 
momentum space.  In the limit that the plasma is assumed to be isotropic, the real 
part of the potential used here reduces to the model originally introduced by Karsch,
Mehr, and Satz (KMS) \cite{Karsch:1987pv} with an additional entropy contribution \cite{Dumitru:2009ni}
and the imaginary part reduces to the result originally obtained by Laine et al \cite{Laine:2006ns}.  
To begin the discussion we first introduce
our ansatz for the one-particle distribution function subject to a momentum-space
anisotropy.

\subsection{The anisotropic plasma}
\label{subsec:aniso}

The phase-space distribution of gluons in the local rest frame is assumed to be given by the
following ansatz~\cite{Dumitru:2007hy,Romatschke:2003ms,
Mrowczynski:2004kv,Romatschke:2004jh,Schenke:2006fz}
\begin{equation}
f({\bf x},{\bf p}) = f_{\rm iso}\left(\sqrt{{\bf p}^2+\xi({\bf p}\cdot{\bf
n})^2 }  / p_{\rm hard}^2 \right) ,  \label{eq:f_aniso}
\end{equation}
where $p_{\rm hard}$ is a scale which specifies the typical momentum
of the particles in the plasma and can be identified with the temperature
in the limit that $\xi=0$.
Thus, $f({\bf x},{\bf p})$ is obtained from an isotropic distribution $f_{\rm
  iso}(|{\bf{p}}|)$ by removing particles with a large momentum
component along ${\bf{n}}$, the direction of anisotropy. In this paper, we will
restrict our consideration to a plasma that is close to equilibrium.   This is
motivated by the fact that in a heavy-ion collision, quarkonium states
are expected to form when the temperature has dropped to (1-2)~$T_c$.
At such temperatures the plasma may have partly equilibrated/isotropized.
Additionally, this means that we can
assume that the function $f_{\rm iso}(|{\bf{p}}|)$ is a thermal
distribution function.

The parameter $\xi$ determines the degree of anisotropy,
\beq
\xi = \frac{1}{2} \frac{\langle {\bf p}_\perp^2\rangle}
{\langle p_z^2\rangle} -1~,
\eeq
where $p_z\equiv \bf{p\cdot n}$ and ${\bf p}_\perp\equiv {\bf{p-n
(p\cdot n)}}$ denote the particle momentum along and perpendicular to
the direction ${\bf n}$ of anisotropy, respectively. If $\xi$ is small,
then it is also related to the shear viscosity of the plasma.  For
example, for one-dimensional boost-invariant expansion governed
by Navier-Stokes evolution
\cite{Asakawa:2006jn,Martinez:2009mf,Martinez:2010sc,Martinez:2010sd}
one finds
\beq \label{eq:xi_eta}
\xi = \frac{10}{T\tau} \frac{\eta}{s}~,
\eeq
where $T$ is the temperature, $\tau$ is the proper time (and $1/\tau$ is
the Hubble expansion rate), and $\eta/s$ is
the ratio of shear viscosity to entropy density. In an expanding
system, non-vanishing viscosity (finite momentum relaxation
rate) implies an anisotropy of the particle momenta which
increases with the expansion rate $1/\tau$. For $\eta/s\simeq
0.1$ -- 0.2 and $\tau T\simeq1$ -- 3 one finds that $\xi\simeq1$.
In general, one can relate $\xi$ to the longitudinal and transverse
pressures in the plasma and it is possible to derive dynamical differential
equations which govern its time evolution similar to viscous hydrodynamics
\cite{Martinez:2010sc,Martinez:2010sd}

We point out that in this paper we restrict ourselves to solving the
time-independent Schr\"odinger equation, i.e.\ we assume that the
plasma is at a constant hard momentum scale $p_{\rm hard}$ and anisotropy $\xi$. This
approximation is useful if the time scale associated with the bound
state, $\sim 1/|E_{\text{bind}}|$, is short compared to the time
scales over which $p_{\rm hard}$ and $\xi$ vary. Indeed, for sufficiently large
quark mass $m_Q$ this condition should be satisfied.

\subsection{The model potential}

Lacking knowledge of the exact heavy-quark potential at finite
temperature, different phenomenological potentials and
lattice-QCD based potentials have been used to
study quarkonium binding energies in the quark-gluon plasma.  
To start, we decompose the potential into real 
and imaginary parts, $V = V_{\rm R} + i V_{\rm I}$. The model for the real part of the potential 
we use was obtained in Ref.~\cite{Dumitru:2009ni}.  The analytic calculation of the 
imaginary part was performed in Refs.~\cite{Laine:2006ns,Laine:2007qy,Dumitru:2009fy}.  The real part is
given by
\beq 
\label{repot}
V_{\rm R}({\bf r}) = -\frac{\alpha}{r} \left(1+\mu \, r\right) \exp\left( -\mu
\, r  \right) + \frac{2\sigma}{\mu}\left[1-\exp\left( -\mu
\, r  \right)\right]
- \sigma \,r\, \exp(-\mu\,r)- \frac{0.8 \, \sigma}{m_Q^2\, r}~,
\eeq
where
\beq
\frac{\mu}{m_D} \equiv 1-\xi \frac{3+\cos 2\theta}{16}~,
\eeq
with $m_D = (1.4)^2 \cdot N_c (1+N_f/6)  \, 4 \pi \alpha_s  \, p_{\rm hard}^2/3$ being the isotropic leading-order Debye mass adjusted
by a factor of $(1.4)^2$ to take into account higher-order corrections \cite{Kaczmarek:2004gv}.  The coupling $\alpha$
folds in a factor of $C_F = (N_c^2 - 1)/(2 N_c)$, i.e. $\alpha \equiv C_F \alpha_s$, where $\alpha_s = g_s^2/(4\pi)$
is the canonically defined strong coupling constant.  We have taken $N_c=3$ and
assumed $N_f=2$ which is appropriate for the temperature range considered herein.
The first term in (\ref{repot}) is a screened Coulomb potential with an entropy addition.  The
second and third terms are a screened linear potential associated with confinement in the low temperature
limit.
The last term in (\ref{repot}) is a relativistic correction which is critical for obtaining accurate
binding energies in the low temperature limit.
For the string tension, we fix $\sigma = 0.223$ GeV and for the strong coupling constant we fix 
$\alpha = 0.385$.\footnote{Since $\alpha_s$ runs logarithmically and therefore has small
variation in the temperature ranges shown, we will ignore the running of the coupling here.  
Incorporating this effect would be straightforward, however, a model of the behavior of $\alpha_s$
at large scales would be required in order fit zero temperature properties of the states considered here.}

The imaginary part is given by \cite{Dumitru:2009fy}
\begin{equation} 
V_{\rm I}({\bf r}) = -\alpha T \biggl[ \phi(\hat{r}) - \xi \left(\psi_1(\hat{r},
\theta)+\psi_2(\hat{r}, \theta)\right)\biggr] ,
\label{impot}
\end{equation}
where $\hat r = m_D r$ and
\begin{eqnarray}
 \phi(\hat{r}) &=& 2\int_0^{\infty}dz \frac{z}{(z^2+1)^2} \left[1-\frac{\sin(z\, \hat{r})}{z\, \hat{r}}\right]~, \\
 \psi_1(\hat{r}, \theta) &=& \int_0^{\infty} dz
 \frac{z}{(z^2+1)^2}\left(1-\frac{3}{2}
 \left[\sin^2\theta\frac{\sin(z\, \hat{r})}{z\, \hat{r}}
 +(1-3\cos^2\theta)G(\hat{r}, z)\right]\right), \\
 \psi_2(\hat{r}, \theta) &=&- \int_0^{\infty} dz
\frac{\frac{4}{3}z}{(z^2+1)^3}\left(1-3 \left[
  \left(\frac{2}{3}-\cos^2\theta \right) \frac
 {\sin(z\, \hat{r})}{z\, \hat{r}}+(1-3\cos^2\theta)
 G(\hat{r},z)\right]\right),\;\;\;\;
\label{funcs}
\end{eqnarray}
with $\theta$ being the angle from the beam direction and
\begin{equation}
 G(\hat{r}, z)= \frac{\hat{r} z\cos(\hat{r} z)- \sin(\hat{r} z)
 }{(\hat{r} z)^3}~.
\label{gdef}
\end{equation}

The short range part of $V_R$ is based on a 
leading order hard-loop perturbation theory calculation presented in Ref.~\cite{Dumitru:2007hy}.
$V_I$ is also obtained from a leading order perturbative calculation \cite{Dumitru:2009fy}.
Being a leading order calculation one may wonder about higher order corrections.
One expects that the leading order calculation pQCD would receive large corrections 
at low temperatures $(T < 10\,T_c)$ since the running coupling becomes large ($g_s > 1$).
For the coupling used above $\alpha_s = 0.29$ one finds $g_s = \sqrt{4 \pi \alpha_s} = 1.9$.
This means that the normal scale hierarchy, $g_s T < T$, implicit in the hard-loop resummation
becomes inverted.\footnote{We note that for temperatures
$T > 2\,T_c$ NNLO perturbative calculations of QCD thermodynamics based on hard-thermal-loop
resummation of QCD agree quite well with available lattice data even though $g_s$ is large 
\cite{Andersen:2010wu,Andersen:2011sf,Andersen:2010ct,Andersen:2009tc}.}
We therefore need to 
supplement the leading order pQCD calculation with a non-perturbative contribution.
For the real part we do this by including a long-range screened linear contribution that is modified
to include an entropy contribution \cite{Dumitru:2007hy}.  In the isotropic limit the resulting
form of the real part potential is in good agreement with lattice data for the heavy quark potential 
\cite{Kaczmarek:2004gv}.  For the imaginary part we currently do not 
have non-perturbative input from lattice calculations with which to constrain the long range 
part; however, we note that calculations of the real and imaginary parts of the potential
using the AdS/CFT correspondence to calculate the corresponding potential in
large t' Hooft coupling limit of ${\cal N}=4$ Supersymmetric Yang-Mills yield similar results 
to those obtained using perturbative QCD \cite{Noronha:2009ia,Noronha:2009da}.
For more information about the relevant scales and limitations of the current approach 
we refer the reader to Sec.~III of Ref.~\cite{Dumitru:2007hy}.

Regarding the length scales which are relevant, we note that the short range part of the potential is appropriate
for describing wavefunctions which have $1/\langle r \rangle < {\cal O} (m_D)$ while the long range
part is relevant if $1/\langle r \rangle > {\cal O} (m_D)$.  Using the form of the real potential
listed above, one finds that the distance scale at which medium effects become large is
roughly given by $r > r_{\rm med} \sim T_c/(2 T)$ fm 
corresponding to $r_{\rm med} \sim 0.25$ fm at 2 $T_c$ \cite{Dumitru:2007hy} .  Numerically, the isotropic Debye mass
used herein is $m_D \sim 3 p_{\rm hard}$, corresponding to $m_D \sim 1.2$ GeV at $p_{\rm hard} = 2 T_c$.  As shown in 
Ref.~\cite{Dumitru:2007hy} Fig.~4, using
the real part of the potential listed above, the RMS radius of the $J/\Psi$ state is approximately
0.8 fm at 2 $T_c$ corresponding to $1/\langle r \rangle \sim$ 250 MeV, which makes the screening
of the long range part of the potential crucially important for fixing the binding energy in this case.  For the
case of the $\Upsilon$ one sees also from Ref.~\cite{Dumitru:2007hy} Fig.~4 that the
RMS radius of the $\Upsilon$ is approximately 0.25 fm corresponding to $1/\langle r \rangle \sim $
800 MeV.  We note importantly that for the $\Upsilon$, due to its relatively small size, the bulk of the medium effect
comes from the temperature dependence of $\lim_{r \rightarrow \infty} V \equiv V_\infty$
(see Fig.~3 of Ref.~\cite{Dumitru:2007hy}).  In closing, one finds that for both the $J/\Psi$ and 
$\Upsilon$ that correct modeling of both the short and long range parts of the potential are
critical for obtaining the temperature dependence of these states.  As mentioned above,
here we extend the results in \cite{Dumitru:2007hy} to include the imaginary part of the potential.
We note that one finds that RMS radii of the states are only weakly affected by inclusion of the
imaginary part of the potential, allowing us to use the estimates above as a rough guide for 
understanding the relevant scales.

\subsection{Analytic estimate in isotropic case}
\label{subsec:est}

In a recent paper \cite{Dumitru:2010id}, Dumitru made an estimate of the effect of
the imaginary part of the potential on the imaginary part of the binding energy of
a quarkonium state.  For this estimate Dumitru assumed a Coulomb wavefunction 
for the quarkonium state and computed the expectation value of the imaginary part 
of the potential exactly in the case of an isotropic plasma.  The result obtained was
\begin{equation}
\label{eq:GammaXi0}
\Gamma(\xi=0) = \frac{T}{\alpha} \frac{m_D^2}{m_Q^2}
\frac{1-(2-\kappa^2)^2 + 4\log\frac{1}{\kappa} }{(1-\kappa^2)^3} ~~~,~~~
\kappa = \frac{1}{\alpha} \frac{m_D}{m_Q}~.
\end{equation}
When plotted in the temperature range between $T_c$ and $3 T_c$ the result above
is approximately linear for both the $J/\psi$ and $\Upsilon$ \cite{Dumitru:2010id}.
For charmonium with $m_Q = 1.3\;{\rm GeV}$ and using the values given for $\alpha$
and $m_D$ in the previous subsection, we obtain a slope consistent with 
$\Gamma \propto (0.08\;{\rm GeV})\,T/T_c$ at $T=0.3$ GeV.  Similarly, 
for bottomonium with $m_Q = 4.7\;{\rm GeV}$ we obtain a slope consistent with $\Gamma \propto 
(0.05\;{\rm GeV})\,T/T_c$.  We note these here for later comparison with numerical 
results presented in the results section.

\section{Numerical Method}
\label{sec:nummethod}

To determine the wavefunctions of bound quarkonium states, we solve
the Schr\"odinger equation
\bea
\hat{H} \phi_\upsilon({\bf x}) &=& E_\upsilon \, \phi_\upsilon({\bf
x})  ~, \nonumber \\
\hat{H} &=& -\frac{\nabla^2}{2 m_R} + V({\bf x}) + m_1 + m_2~,
\label{3dSchrodingerEQ}
\eea
on a three-dimensional lattice in coordinate space with the potential
given by $V = V_{\rm R} + i V_{\rm I}$ where the real and imaginary
parts are specified in Eqs.~(\ref{repot}) and (\ref{impot}), respectively.  
Here, $m_1$ and $m_2$ are the masses
of the two heavy quarks and $m_R$ is the reduced mass, $m_R = m_1
m_2/(m_1+m_2)$. The index $\upsilon$ on the eigenfunctions,
$\phi_\upsilon$, and energies, $E_\upsilon$, represents a list of all
relevant quantum numbers, such as\ $n$, $l$, and $m$ for a radial Coloumb
potential. Due to the anisotropic screening scale, the wavefunctions
are no longer radially symmetric if $\xi \neq 0$. Since we consider
only small anisotropies we nevertheless label the states as $1S$
(ground state) and $1P$ (first excited state), respectively.

To find solutions to Eq.~(\ref{3dSchrodingerEQ}), we use the finite
difference time domain method (FDTD)~\cite{Sudiarta:2007,Strickland:2009ft}.  In this
method we start with the time-dependent Schr\"odinger equation
\beq
i \frac{\partial}{\partial t} \psi({\bf x},t) = \hat H \psi({\bf x},t) \, ,
\label{3dSchrodingerEQminkowski}
\eeq
which can be solved by expanding in terms of the eigenfunctions,
$\phi_\upsilon$:
\beq \psi({\bf x},t) = \sum_\upsilon c_\upsilon \phi_\upsilon({\bf x})
e^{- i E_\upsilon t}~.
\label{eigenfunctionExpansionMinkowski}
\eeq
If one is only interested in the lowest energy states (ground state
and first few excited states) an efficient way to proceed is to
transform~(\ref{3dSchrodingerEQminkowski})
and~(\ref{eigenfunctionExpansionMinkowski}) to Euclidean time using a
Wick rotation, $\tau \equiv i t$:
\beq \frac{\partial}{\partial \tau} \psi({\bf x},\tau) = - \hat H
\psi({\bf x},\tau) \, ,
\label{3dSchrodingerEQeuclidean}
\eeq
and
\beq \psi({\bf x},\tau) = \sum_\upsilon c_\upsilon \phi_\upsilon({\bf
x}) e^{- E_\upsilon \tau} ~.
\label{eigenfunctionExpansionEuclidean}
\eeq
For details of the discretizations used etc. we refer the reader 
to Refs.~\cite{Strickland:2009ft,Sudiarta:2007}.

\subsection{Finding the ground state}

By definition, the ground state is the state with the lowest energy
eigenvalue, $E_0$. Therefore, at late imaginary time the sum
over eigenfunctions (\ref{eigenfunctionExpansionEuclidean}) is
dominated by the ground state eigenfunction
\beq \lim_{\tau \rightarrow \infty} \psi({\bf x},\tau) \rightarrow c_0
\phi_0({\bf x}) e^{- E_0 \tau}~.
\label{groundstateEuclideanLateTime}
\eeq
Due to this, one can obtain the ground state wavefunction,
$\phi_0$, and energy, $E_0$, by solving
Eq.~(\ref{3dSchrodingerEQeuclidean}) starting from a random
three-dimensional wavefunction, $\psi_{\text{initial}}({\bf x},0)$,
and evolving forward in imaginary time. This initial wavefunction
should have a nonzero overlap with all eigenfunctions of the
Hamiltonian; however, due to the damping of higher-energy
eigenfunctions at sufficiently late imaginary times we are left with
only the ground state, $\phi_0({\bf x})$. Once the ground state
wavefunction (or any other wavefunction) is found, we can
compute its energy eigenvalue via
\bea
E_\upsilon(\tau\to\infty) = \frac{\langle \phi_\upsilon | \hat{H} |
\phi_\upsilon \rangle}{\langle \phi_\upsilon | \phi_\upsilon
\rangle} = \frac{\int d^3{\bf x} \, \phi_\upsilon^*
\, \hat{H} \, \phi_\upsilon }{\int d^3{\bf x} \, \phi_\upsilon^*
\phi_\upsilon} \; .
\label{bsenergy}
\eea

To obtain the binding energy of a state,
$E_{\upsilon,\text{bind}}$, we subtract the quark masses and
the real part of the potential at infinity
\beq
E_{\upsilon,\text{bind}} \equiv E_\upsilon - m_1 - m_2 -
\frac{\langle \phi_\upsilon | {\rm Re}[V(\theta,|{\bf r}|\to\infty)] | \phi_\upsilon
\rangle}{\langle \phi_\upsilon | \phi_\upsilon \rangle} \; .
\label{bsbindingenergy}
\eeq
For the isotropic KMS potential the last term is independent of the
quantum numbers $\upsilon$ and equal to $2\sigma/m_D$. In the
anisotropic case, however, this is no longer true since the operator
$V_\infty(\theta)$ carries angular dependence, as discussed
above. Its expectation value is, of course, independent of $\theta$ but
does depend on the anisotropy parameter $\xi$.

\subsection{Finding the excited states}

The basic method for finding excited states is to first evolve the
initially random wavefunction to large imaginary times, find the
ground state wavefunction, $\phi_0$, and then project this state out
from the initial wavefunction and re-evolve the partial-differential
equation in imaginary time. However, there are (at least) two more
efficient ways to accomplish this. The first is to record snapshots of
the 3d wavefunction at a specified interval $\tau_{\text{snapshot}}$
during a single evolution in $\tau$. After having obtained the ground
state wavefunction, one can go back and extract the excited
states by projecting out the ground state wavefunction from the
recorded snapshots of $\psi({\bf x},\tau)$.

An alternative way to select different excited states is to impose a
symmetry condition on the initially random wavefunction which cannot
be broken by the Hamiltonian evolution. For example, one can select
the first excited state of the (anisotropic) potential by
anti-symmetrizing the initial wavefunction around either the $x$, $y$,
or $z$ axes.  In the anisotropic case this trick can be used to
separate the different polarizations of the first excited state of the
quarkonium system and to determine their energy eigenvalues with high
precision.  This high precision allows one to more accurately
determine the splitting between polarization states which are
otherwise degenerate in the isotropic Debye-Coulomb potential.

Whichever method is used, once the wavefunction of an excited state
has been determined one can again use the general
formulas~(\ref{bsenergy}) and~(\ref{bsbindingenergy}) to determine 
the excited state binding energy.  For code benchmarks and tests see 
App.~\ref{app:bench}.
\begin{figure}[t]
\vspace{1mm}
\includegraphics[width=16cm]{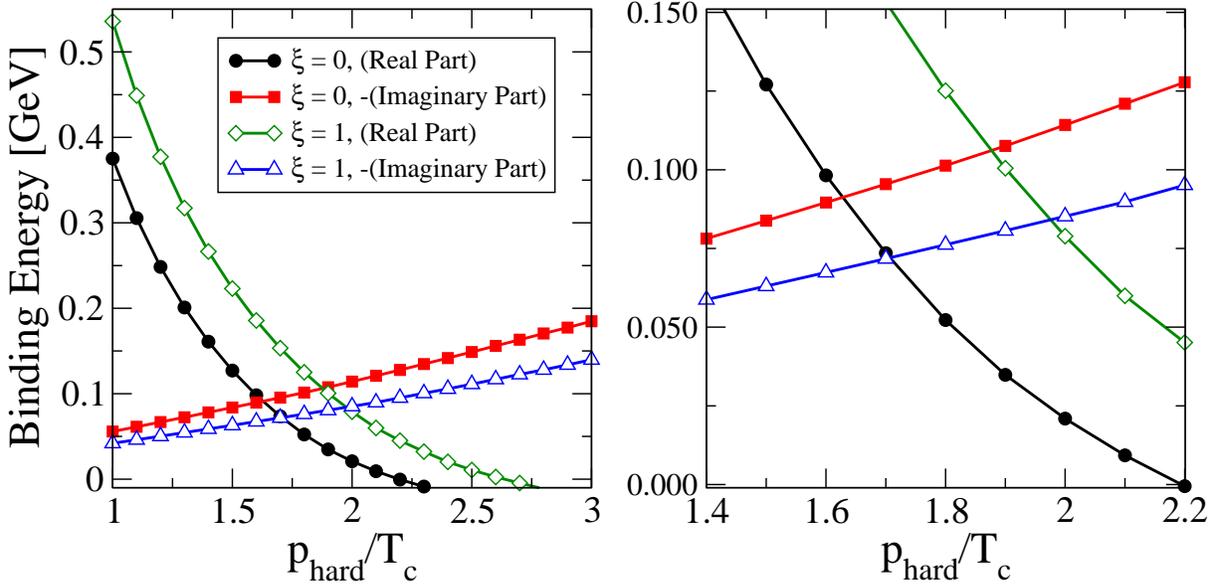}
\caption{Real and imaginary parts of the charmonium ground state ($J/\psi$) binding energy as a function of $p_{\rm hard}$.
Both isotropic $\xi=0$ and anisotropic $\xi=1$ cases are shown.  The left panel shows full temperature range and the right
panel focuses on the region where the real and imaginary parts become comparable.  
See text for parameters such
as lattice size, lattice spacing, etc.}
\label{fig:charmonium}
\end{figure}

\begin{figure}[t]
\vspace{1mm}
\includegraphics[width=11cm]{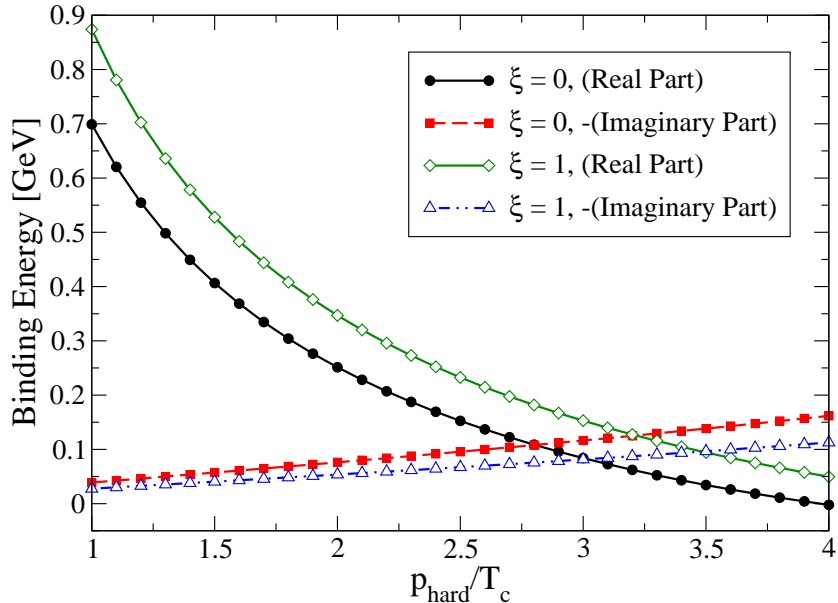}
\caption{Real and imaginary parts of the bottomonium ground state ($\Upsilon$) binding energy as a function of $p_{\rm hard}$.
Both isotropic $\xi=0$ and anisotropic $\xi=1$ cases are shown.  See text for parameters such
as lattice size, lattice spacing, etc.}
\label{fig:bottomonium}
\end{figure}

\begin{figure}[t]
\vspace{1mm}
\includegraphics[width=11cm]{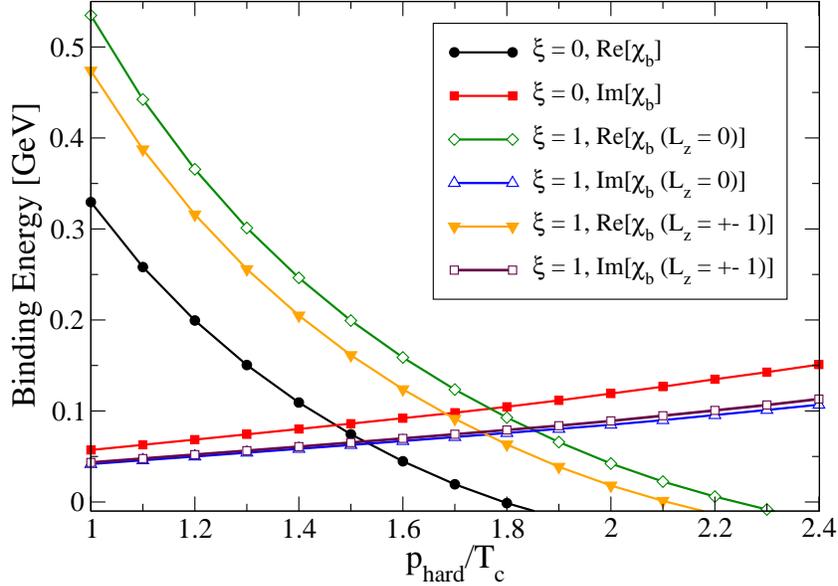}
\caption{Real and imaginary parts of $\chi_b$ binding energy as a function of $p_{\rm hard}$.
Both isotropic $\xi=0$ and anisotropic $\xi=1$ cases are shown.  See text for parameters such
as lattice size, lattice spacing, etc.}
\label{fig:chib}
\end{figure}

\section{Results and Discussion}
\label{sec:results}

In this section we present results for isotropic ($\xi=0$) and anisotropic ($\xi=1$) binding
energies for charmonium ($J/\psi$), bottomonium ($\Upsilon$), and the first excited state of 
bottomonium ($\chi_b$) as a function of the hard momentum scale $p_{\rm hard}$.  
We will first assume that $p_{\rm hard}$ is held constant and vary the anisotropy parameter.
Note increasing $\xi$ results in a 
decrease in the density since $n \propto p_{\rm hard}^3/\sqrt{1+\xi}$ \cite{Dumitru:2007hy}.  
This reduced density results in less
Debye screening and thus a more strongly bound state.  We therefore expect that states
with large anisotropy will have increased binding energies compared to the isotropic states. 
One could imagine holding another thermodynamic property such as the number density or energy 
density constant as one changes the anisotropy parameter.  We will return to this issue at the end of 
this section and show that the results in these cases can be obtained from a simple rescaling of the results
presented below.  In all plots shown, we assume $T_c = 192$ MeV and fix the imaginary-time step
in the numerical algorithm to be $\Delta\tau = a^2/8$ where $a$ is the spatial lattice spacing.

\subsection{Results as a function of the hard momentum scale}
\label{sec:fixedphard}

In Fig.~\ref{fig:charmonium} we plot the binding energy of the charmonium ground state ($J/\psi$)
as a function of $p_{\rm hard}$.  For this figure, we used a lattice 
size of $256^3$ with lattice dimension of $L = 25.6$ GeV$^{-1}$ and a lattice spacing of
$a = 0.1$  GeV$^{-1}$.  For the charmonium mass, we used $m_c = 1.3$ GeV.
In Fig.~\ref{fig:charmonium} we show both the real part (black
line with filled circles) and the imaginary part (red line with filled squares) of the
isotropic ground state binding energy.  Comparing these two curves, we see that the imaginary part of the binding energy 
becomes comparable to the real part at $p_{\rm hard} \sim 1.63\,T_c$.  In contrast, in the anisotropic case ($\xi=1$)
we find that the intersection between the imaginary (blue line with open triangles)
and real parts (green line with open diamonds) occurs at
$p_{\rm hard} \sim 1.88\,T_c$.  In the range between 1 and 3 $T_c$
we obtain a slope of $4.9\times10^{-2}$ GeV for the imaginary part of the binding energy when
$\xi=0$ and $6.4\times10^{-2}$ GeV when $\xi=1$.  In the isotropic case Dumitru's 
perturbative calculation \cite{Dumitru:2010id} gives a slope of $8\times10^{-2}$ GeV.
Our method is non-perturbative since we don't assume perturbations around Coulomb 
wave functions, so one should not
be surprised to see some important differences.

In Fig.~\ref{fig:bottomonium} we plot the binding energy of the
bottomonium ground state ($\Upsilon$) as a function of $p_{\rm hard}$.  
For this figure, we used a lattice 
size of $256^3$ with lattice dimension of $L = 25.6$ GeV$^{-1}$ and a lattice spacing of
$a = 0.1$  GeV$^{-1}$.  For the bottomonium mass, we used $m_b = 4.7$ GeV.
In Fig.~\ref{fig:bottomonium} we show both the real part (black
line with filled circles) and the imaginary part (red line with filled squares) of
the isotropic ground state binding energy.  When $\xi=0$, we see that the imaginary part of the binding energy 
becomes comparable to the real part
at $p_{\rm hard} \sim 2.8\,T_c$.  In the anisotropic case ($\xi=1$)
we find that the intersection between the imaginary (blue line with open triangles)
and real parts (green line with open diamonds) occurs at approximately
$3.5\,T_c$.  For $\xi=0$, in the range between 1 and 4 $T_c$
we obtain a slope of $2.8\times10^{-2}$ GeV for the imaginary part of the binding energy.
In the anisotropic case ($\xi=1$) we find a slope of $4.2\times10^{-2}$ GeV.  We
can once again compare to the analytic result of Dumitru \cite{Dumitru:2010id}
which gives an isotropic slope of $5\times10^{-2}$ for the $\Upsilon$.  Once
again, the numbers are roughly in agreement.

In Fig.~\ref{fig:chib} we plot the binding energy of the first p-wave excited state of 
bottomonium ($\chi_b$) as a function of $p_{\rm hard}$.  For this figure we used a lattice 
size of $256^3$ with lattice dimension of $L =38.4$ GeV$^{-1}$ and a lattice spacing of
$a = 0.15$  GeV$^{-1}$.  For the bottomonium mass, we used $m_b = 4.7$ GeV.
As was the case with the bottomonium ground state we see an increase in the real 
part of the binding energy with increasing anisotropy.
Most importantly, we find that there is an approximately 60 MeV
splitting between the $L_z = 0$ and $L_z = \pm 1$ states with the states with 
$L_z = \pm 1$ having the lower binding energy.  We would therefore expect fewer
$L_z = \pm 1$ states of the $\chi_b$ to be produced in an anisotropic plasma.  Determining 
precisely how many fewer would be produced requires knowledge of the time evolution of the 
momentum scale $p_{\rm hard}$ and anisotropy $\xi$.

\subsection{Fixing number density or energy density}

\begin{figure}[t]
\vspace{1mm}
\includegraphics[width=11cm]{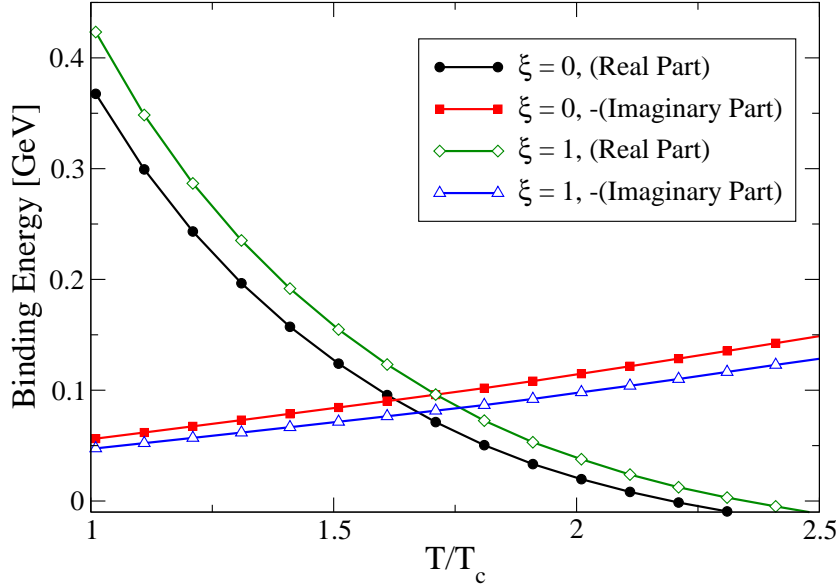}
\caption{Real and imaginary parts of the charmonium ground state ($J/\psi$) binding energy 
as a function of temperature assuming fixed number density.
Both isotropic $\xi=0$ and anisotropic $\xi=1$ cases are shown. 
See text for parameters such
as lattice size, lattice spacing, etc.}
\label{fig:charmonium-nd}
\end{figure}

\begin{figure}[t]
\vspace{1mm}
\includegraphics[width=11cm]{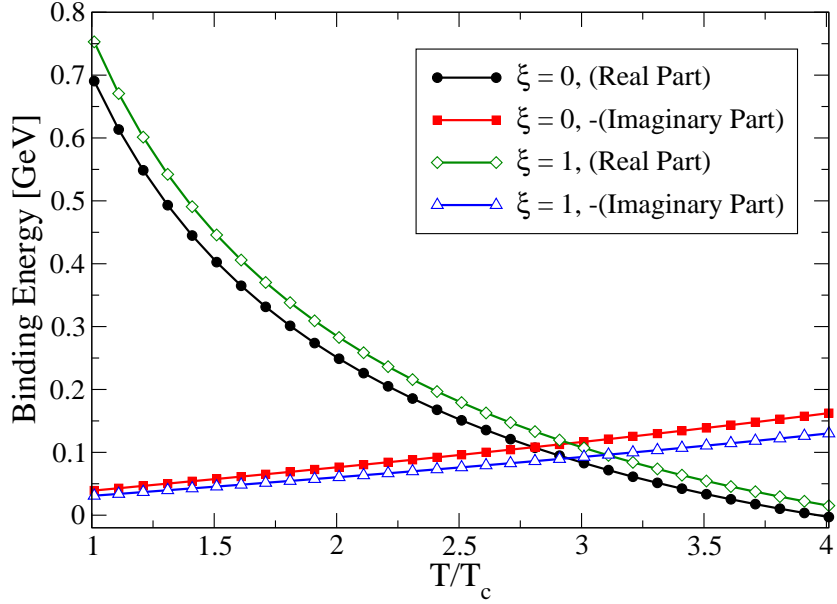}
\caption{Real and imaginary parts of bottomonium ground state ($\Upsilon$) binding energy as a function of temperature
assuming fixed number density.
Both isotropic $\xi=0$ and anisotropic $\xi=1$ cases are shown.  See text for parameters such
as lattice size, lattice spacing, etc.}
\label{fig:bottomonium-nd}
\end{figure}

\begin{figure}[t]
\vspace{1mm}
\includegraphics[width=11cm]{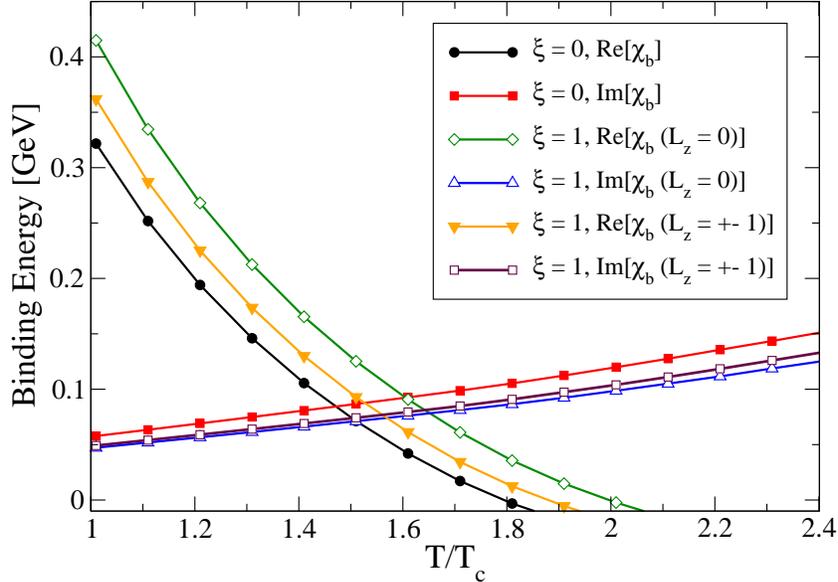}
\caption{Real and imaginary parts of the $\chi_b$ binding energy as a function of temperature
assuming fixed number density.
Both isotropic $\xi=0$ and anisotropic $\xi=1$ cases are shown.  See text for parameters such
as lattice size, lattice spacing, etc.}
\label{fig:chib-nd}
\end{figure}

As mentioned in the beginning of this section, when one is working in a non-equilibrium setting
it is necessary to specify which quantities are held fixed.  In equilibrium, it is sufficient
to specify the temperature.  The temperature then uniquely determines the number density, energy
density, etc.  In the previous subsection we presented results obtained when one
holds the hard momentum scale $p_{\rm hard}$ fixed while varying the anisotropy
parameter $\xi$.  Doing so, however, results in different number densities and energy
densities for different anisotropies ($\xi$).  Here we discuss how to fix either the number
or energy density by adjusting $p_{\rm hard}$ appropriately.  We first demonstrate this in the case
of the number density and show that for small anisotropy the scalings required
to fix the number density or energy density are practically identical.  We then present results
for the binding energies of the states we are interested in for the case of fixed number
density,  since in this paper we concentrate on anisotropies which are small enough
that the difference between the cases of fixed number density and fixed energy density
is numerically very small.

The number density as a function of $\xi$ and $p_{\rm hard}$
can be calculated for an arbitrary isotropic distribution $f_{\rm iso}$
\cite{Romatschke:2004jh}
\beq
n(\xi,p_{\rm hard}) = \frac{n_{\rm iso}(p_{\rm hard})}{\sqrt{1+\xi}} \, ,
\label{eq:number}
\eeq
where $n_{\rm iso}$ is the number density associated with the isotropic
distribution function $f_{\rm iso}$ via
\beq
n_{\rm iso}(p_{\rm hard}) = \int  \frac{d^3p}{(2\pi)^3}   f_{\rm iso}(|{\bf p}|,p_{\rm hard}) \, .
\eeq
Since $n_{\rm iso}$ only contains one dimensionful scale, by dimensional analysis we have 
$n_{\rm iso} \propto p_{\rm hard}^3$.  In order to keep the number density 
(\ref{eq:number}) fixed as one changes $\xi$, one can adjust $p_{\rm hard}$ by requiring
\beq
p_{\rm hard} = (1+\xi)^{1/6} \, T  \;\;\;\;\; {\rm[fixed\;number\;density]} \; ,
\label{eq:scalen}
\eeq
where $T$ is the corresponding isotropic scale (temperature) which gives the target
number density when $\xi=0$, i.e. $n_{\rm iso}(T)$.

Similarly, the energy density as a function of $\xi$ and $p_{\rm hard}$
can be calculated for an arbitrary isotropic distribution $f_{\rm iso}$ \cite{Martinez:2008di}
\beq
{\cal E}(\xi,p_{\rm hard}) = {\cal R}(\xi) {\cal E}_{\rm iso}(p_{\rm hard}) \, ,
\label{eq:energy}
\eeq
where ${\cal E}_{\rm iso}$ is the energy density associated with the isotropic
distribution function $f_{\rm iso}$ and
\beq
{\cal R}(\xi) = \frac{1}{2}\left(\frac{1}{1+\xi}
+\frac{\arctan\sqrt{\xi}}{\sqrt{\xi}} \right) \, .
\eeq
Since ${\cal E}_{\rm iso}$ only contains one dimensionful scale, by dimensional analysis
we have ${\cal E}_{\rm iso} \propto p_{\rm hard}^4$ and we can fix the energy density
to the corresponding isotropic energy density with scale $T$ by requiring
\beq
p_{\rm hard} =T/ [{\cal R}(\xi)]^{1/4} \;\;\;\;\; {\rm[fixed\;energy\;density]} \; .
\label{eq:scalee}
\eeq

The scalings for fixed number density (\ref{eq:scalen}) and fixed energy density (\ref{eq:scalee})
are different; however, in the limit of small anisotropies the scalings are very close.
Expanding to quadratic order, one finds
\begin{subequations}
\begin{align}
\label{eq:nsxi}
\frac{p_{\rm hard}}{T} &=  1+\frac{1}{6}\xi - \frac{29}{360} \xi^2 + {\cal O}(\xi^3)    &{\rm[fixed\;number\;density]} \; , \\
\label{eq:esxi}
\frac{p_{\rm hard}}{T} &=  1+\frac{1}{6}\xi - \frac{5}{72} \xi^2 + {\cal O}(\xi^3)        &{\rm[fixed\;energy\;density]} \; ,
\end{align}
\end{subequations}
which agree at linear order and differ by 7.4\% in the quadratic coefficient.  One finds that, when
including all orders in the expansion, the right hand sides of (\ref{eq:nsxi}) and (\ref{eq:esxi})
differ by only 0.25\% at $\xi=1$.  Therefore, for the range of anisotropies considered here, the two scalings are
functionally equivalent.  We will therefore only present results for fixed number density with the 
understanding that the fixed energy density results are indistinguishable by the human eye.

In Figs.~\ref{fig:charmonium-nd}, \ref{fig:bottomonium-nd}, and \ref{fig:chib-nd}, we show
the binding energies which result from the fixed number density rescaling of the horizontal axes of
Figs.~\ref{fig:charmonium}, \ref{fig:bottomonium}, and \ref{fig:chib}.  As can be seen from
Figs.~\ref{fig:charmonium-nd}, \ref{fig:bottomonium-nd}, and \ref{fig:chib-nd}, requiring
fixed number/energy density weakens the effect of anisotropies on the ground state binding
energies.  In the case of the ground states of charmonium and bottomonium shown in
Figs.~\ref{fig:charmonium-nd} and \ref{fig:bottomonium-nd} we find that the splitting
between the $\xi=0$ and $\xi=1$ cases at the critical temperature is approximately 50 MeV
in both cases.  

Finally, we emphasize that in the case of the first excited states of bottomonium shown in 
Fig.~\ref{fig:chib-nd} the splitting between the $L_z=0$ and $L_z=\pm1$ states is unaffected
by the rescaling since we have $\xi=1$ for both states.  Therefore, one has a relatively clean observable
that is sensitive to plasma anisotropies regardless of the quantity which is assumed to be 
held fixed.

\section{Conclusions}
\label{sec:conc}

In this paper we have presented first results on the effect of including both the real and imaginary parts
of the heavy quarkonium potential on the binding energies of the charmonium ground state ($J/\psi$), 
the bottomonium ground state ($\Upsilon$), and the first $p$-wave excited state of bottomonium ($\chi_b$).  
We did this by numerically solving the 
three-dimensional Schr\"odinger equation for the complex potential given by Eqs.~(\ref{repot})
and (\ref{impot}).  This enabled us to extract both the real and imaginary parts of the binding
energies for the states.  Using our model potential, we investigated both isotropic and weakly 
anisotropic plasmas.  We found that, there can be a sizable effect of momentum-space
anisotropy on both the real and imaginary parts of the quarkonium binding energy.  One can 
estimate the disassociation temperature of the states by determining the temperature at which
the real and imaginary parts of the binding energy become the same.  Using this criteria,
in the isotropic case we estimate the $J/\psi$, $\Upsilon$ and $\chi_b$ to have disassociation 
temperatures of 1.6 $T_c$, 2.8 $T_c$, and 1.5 $T_c$, respectively.  We note, however,
that even prior to these disassociation temperatures the states will be suppressed
due to the exponential decay of the states with a rate related to the imaginary part
of the binding energy.  We plan to investigate the phenomenological impact of our results
on the time evolution of quarkonium decay in a future publication.

In the case of a plasma with a finite momentum-space anisotropy, we presented results
for both fixed hard momentum scale and fixed number density. Our results demonstrate that
the corresponding anisotropic states have a higher binding energy in accordance with previous results that
employed only the real part of the quarkonium potential used herein \cite{Dumitru:2009ni}.  
We showed that, for small anisotropy, fixing the number density and fixing the energy
density gives results which are the same to within less than a fraction of a percent.  
We demonstrated that fixing the number density reduces the effect of anisotropy compared
to the case of fixing the hard momentum scale, but does not completely remove the effect
of momentum-space anisotropy on the binding energies.  Finally, we emphasized the importance
of the finite-anisotropy splitting between the $\chi_b$ states with $L_z =0$ and $L_z=\pm1$. 
This splitting is  independent of whether one fixes the hard momentum scale, number density, or 
energy density.  Therefore, this splitting represents a possible observable which could be used
to determine the time-averaged plasma anisotropy parameter.

Looking forward, to fully assess the phenomenological impact of plasma momentum-space
anisotropies on quarkonium states requires the convolution of the results presented here 
with the space-time evolution
of the hard momentum scale and anisotropy parameter.  A method for determining the 
dynamical evolution of these parameters has recently been determined \cite{Martinez:2010sc,%
Martinez:2010sd}.  In addition, since these works show that $\xi$ can become large, it
will be necessary to investigate the effect of large anisotropies on quarkonium binding
energies.  The calculations necessary to address these questions are currently underway.

\section*{Acknowledgments}

We thank A.~Dumitru for discussions.  K. McCarty was supported during the summer of 2010
by the Cormack Fund.  M. Strickland was supported in part by the Helmholtz 
International Center for FAIR Landesoffensive zur Entwicklung Wissenschaftlich-\"Okonomischer
Exzellenz program.

\section*{Note added}

In arXiv versions 1-3 there was a mistake in the final results for the imaginary part
of the binding energies.  The mistake stems from the fact that we had 
subtracted the full complex-valued potential at infinity $V_\infty$; however,
formally only the real part of $V_\infty$ should be subtracted since the imaginary part of 
$V_\infty$ is related to heavy quark damping in the plasma which is physically relevant.  As a
consequence all imaginary parts of the binding energies are changed and we have updated
all figures.  While the results are qualitatively similar, the key change is that the imaginary
part of the binding energy now has a stronger dependence on the anisotropy parameter, 
$\xi$, in most cases.

\appendix

\section{Numerical Tests}
\label{app:bench}

\subsection{Convergence Test}

\begin{figure}[t]
\includegraphics[width=16cm]{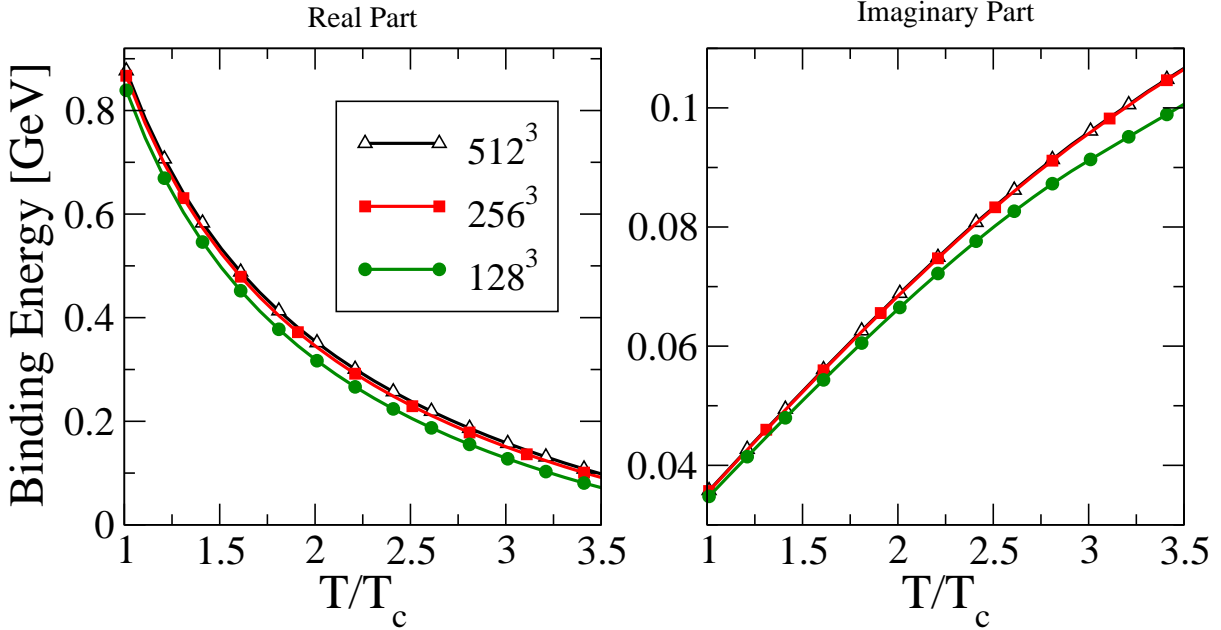}
\caption{Real and imaginary parts of the bottomonium binding energy for three different 
lattice sizes of $128^3, 256^3$, and $512^3$.}
\label{fig:bottomonium-convergence}
\end{figure}

In this appendix we present some convergence data for a particular state in order to
demonstrate the approach to the continuum limit.
In Fig.~\ref{fig:bottomonium-convergence} we show both the real and imaginary parts of the
bottomonium ground state
binding energy for three different lattice spaces of $128^3, 256^3$, and $512^3$.
For each of the runs, the lattice size was fixed to $L = 25.6$ GeV$^{-1}$ with the lattice spacing
in each case given by $a=0.2$, $a=0.1$, and $a=0.05$ GeV$^{-1}$, respectively.  We chose
$\xi=1$, $m_b = 4.7$ GeV and used an imaginary-time step given by 
$\Delta \tau = a^2/8$ in each case.

As can be seen 
from this figure, there is a larger effect due to reducing the lattice space 
on the real part than the imaginary part.  This is to
be expected since the real part contains a divergence at the origin, whereas the imaginary 
part is regular.  In the case of the real part at 3 $T_c$, we see an approximately 8.0\% 
correction when going from $128^3$ to $256^3$ and a 2.6\% correction when going from $256^3$ 
to $512^3$.  The corrections in the case of the imaginary part are 2.4\% and
0.16\%, respectively.  Therefore we see that the $256^3$ runs presented in the body
of the text are reliable up to corrections on the order of 3\%.

\subsection{Harmonic Oscillator with complex spring constant}

In this part of the appendix we explore the ability of the FDTD algorithm to handle the case of 
complex potentials.  We investigate a simple
one dimensional test case which consists of solving the Schr\"{o}dinger equation for a
particle of mass $m$ in a harmonic oscillator potential which has a complex spring
constant $k$.  We first derive the analytic solution and then compare the output of the
code and the analytic solution.

Our goal is to solve the time-independent Schr\"{o}dinger equation for a particle of mass $m$ 
which is bounded by the quadratic potential $V(x) = k x^2/2$, in the case that $k$ is complex.  
Before proceeding, we review the solution in the case
that $k$ is real, writing it first in terms of Parabolic Cylinder Functions and then 
showing how these reduce to the Hermite Polynomials.

\subsubsection{Review of the case of a real spring constant}

We are interested in solving the time independent 1D Schr\"{o}dinger equation for the harmonic oscillator
potential
\begin{equation}
-\frac{\hbar^2}{2m}\frac{d^2 \psi}{d x^2} + \frac{k}{2}x^2\psi = E\psi \,.
\label{schr1}
\end{equation}
To begin with, we introduce the variables
$\omega \equiv \sqrt{\frac{k}{m}}$,
$\alpha \equiv \left(\frac{mk}{\hbar^2}\right)^{1/4}$,
$\lambda \equiv \frac{2E}{\hbar\omega}$,
and $u \equiv \alpha x$ which allow us to write (\ref{schr1}) compactly as
\begin{equation}
\frac{d^2 \psi}{d u^2} - (u^2 - \lambda) \psi = 0 \, .
\label{schr4}
\end{equation}

\subsubsection{Asymptotic Behavior}

We now want to find solutions for our wavefunction at $|u|\rightarrow\infty$.  
In the limit $|u|>>1$ Eq.~(\ref{schr4}) becomes
\begin{equation}
\lim_{|u|\rightarrow \infty} \frac{d^2 \psi}{d u^2} = u^2 \psi \, ,
\end{equation}
which has an asymptotic solutions of the form
\begin{equation}
\lim_{|u|\rightarrow \infty} \psi(u) = A\,u^p\,e^{-u^2/2}   \, ,
\end{equation}
or
\begin{equation}
\lim_{|u|\rightarrow \infty} \psi(u) = B\,u^q\,e^{u^2/2} \, ,
\end{equation}
with A and B being arbitrary constants.  Since $\psi$ must remain finite as $|u|\rightarrow \infty$, this
requires that we discard the second solution.  Requiring that the wavefunction be single valued for negative $u$
implies that $p$ must be an integer.\footnote{Note that one could move the cut along the negative $u$ axis into the complex plane, so
the more properly-stated requirement is that the wavefunction be single-valued everywhere in the complex plane.}  This yields an asymptotic solution of the form
\begin{equation}
\lim_{|u|\rightarrow \infty} \psi(u)= A\,u^p e^{-u^2/2}\, ,
\label{asymptoticform}
\end{equation}
where $p$ is integer-valued.

\subsubsection{Solution in terms of Parabolic Cylinder Functions}

We now define new variables $a = -\lambda/2$ and 
$b = \sqrt{2} u$ such  that $u^2 = b^2/2$ and $2 \, d^2/ d b^2 = d^2/ d u^2$.
Substituting these into Eq. (\ref{schr4}) gives
\begin{equation}
\frac{d^2 \psi}{d b^2} - \left(\frac{1}{4} b^2 + a \right) \psi = 0 \, .
\end{equation}
The solution to this differential equation is given by the Parabolic Cylinder Function D from Chapter 
19 of Abramowitz and Stegun \cite{abramowitz+stegun} $\psi (a,b) = U(a,b) = 
D_{-a-\frac{1}{2} b}(b)$.  If $U(a,b)$ is a solution, then so are $U(a,-b), U(-a,ib)$, and  
$U(-a,-ib)$.  This leaves us with the general solution
\begin{equation}
\psi (u) = A\,U\left(-\frac{\lambda}{2}, \sqrt{2} u \right) + B\,U\left(\frac{\lambda}{2}, i\sqrt{2} u \right) .
\label{gensoln}
\end{equation}

\subsubsection{Matching Asymptotic Behavior}

Now we analyze the asymptotic behavior of $U$ when $b \gg a$.  For $b \gg a$, we 
have \cite{abramowitz+stegun}
\begin{equation}
U(a,b) \sim e^{-\frac{1}{4} b^2} b^{-a-\frac{1}{2}}(1 - {\cal O}(b^{-2})) .
\end{equation}
We see that in the limit $b \rightarrow \infty$ that $U(a,ib)$ approaches $\mbox{+}\infty$.
Since the wavefunction should be finite at  $b \rightarrow \infty$ this requires $B=0$ in 
Eq. (\ref{gensoln}).
We can solve for $\lambda$ by matching to the asymptotic form given in 
Eq.~(\ref{asymptoticform})
\begin{equation}
\left(-\frac{\lambda}{2}+\frac{1}{2}\right) = -p .
\end{equation}
where $p$ must be an integer as discussed previously.  Solving for $\lambda$ gives us
$\lambda = 2p + 1$,
which tells us that $\lambda$ is discrete.  This leaves us with the solution
to the quantum harmonic oscillator differential equation 
in terms of Parabolic Cylinder D functions
\begin{equation}
\psi (u) = A\,U\!\left(-\frac{2p+1}{2},\sqrt{2} u \right) .
\label{soln1}
\end{equation}

\subsubsection{Connection to Hermite Polynomials}

When $n$ is a non-negative integer, $U(-n-1/2,b)$ is expressible in terms of Hermite 
Polynomials~\cite{abramowitz+stegun}
\begin{equation}
U(-n-1/2,b) = 2^{-\frac{1}{2} n}e^{-\frac{1}{4} b^2}H_n\left(\frac{b}{\sqrt{2}}\right) ,
\end{equation}
where $H_n(x)$ is a Hermite polynomial.  Using this we find that 
Eq. (\ref{soln1}) can be expressed as
\begin{equation}
\psi_p(u) = A_p\,e^{-\frac{1}{2} u^2}\,H_p(u) .
\end{equation}
This is the standard textbook form of the quantum harmonic oscillator wavefunctions.
We now extend the general solution (\ref{gensoln}) to the case where $k$ can be 
complex.

\subsection{Extending the solution to the case of a complex spring constant}

We now examine the behavior of our solution when the spring constant $k$ is complex.
To begin, we note a symmetry of Eq.~(\ref{schr4}): namely, if we find that a solution
for a given $k$ is given by $\psi$, then the solution for the complex conjugate spring
constant $k^*$ will be given by $\psi^*$.  Therefore, it suffices to solve the equation
in only half of the complex $k$ plane.  In order to find the specific solution, we take the 
general solution in Eq. (\ref{gensoln})
and re-examine its asymptotic behavior.  Depending on the angle of $k$ in the complex plane,
we find that one needs to set either $A$ or $B$ in Eq.~(\ref{gensoln}) to zero. For large values of 
$u>>|a|$ we have from Ref.~\cite{abramowitz+stegun}
\begin{equation}
U(a,u) \sim e^{-\frac{1}{4}u^2}u^{-a-\frac{1}{2}} ,
\end{equation}
where we have dropped a factor of $\sqrt{2}$ for simplicity.
For complex-valued $u=\beta+i\gamma$ with real-valued $\beta$ and $\gamma$, we have three cases 
that determine the asymptotic convergence of $U(a,u)$:  $\beta>\gamma$, $\beta<\gamma$, and 
$\beta=\gamma$.

\subsubsection{Case A: Real part greater than imaginary part ($\beta>\gamma$) }

In this case, we must require $B=0$ in Eq. (\ref{gensoln}) and we have
\begin{eqnarray}
\psi(u)=U(a,\beta+i\gamma)&\thicksim& e^{-\frac{1}{4}(\beta+i\gamma)^2}u^{-a-\frac{1}{2}}\\
&=&e^{-\frac{1}{4}(\beta^2+2i\beta\gamma-\gamma^2)}u^{-a-\frac{1}{2}}\nonumber\\
&=&e^{-\frac{1}{4}\beta^2}e^{-\frac{1}{2}i\beta\gamma}e^{\frac{1}{4}\gamma^2}u^{-a-\frac{1}{2}}\nonumber ,
\end{eqnarray}
which converges since $\beta>\gamma$.\\

\subsubsection{Case B: Real part less than imaginary part ($\beta<\gamma$)}

In this case, we require $A=0$ in Eq. (\ref{gensoln}) and we have
\begin{eqnarray}
U(-a,iu)&=&U(-a,i \beta-\gamma)\\
&\thicksim& e^{-\frac{1}{4}(i \beta-\gamma)^2}u^{a-\frac{1}{2}}\nonumber\\
&=& e^{-\frac{1}{4}(-\beta^2-2i\beta\gamma+\gamma^2)}u^{a-\frac{1}{2}}\nonumber\\
&=& e^{\frac{1}{4}\beta^2}e^{+\frac{1}{2}i\beta\gamma}e^{-\frac{1}{4}\gamma^2}u^{a-\frac{1}{2}}\nonumber ,
\end{eqnarray}
which converges since $\beta<\gamma$.

\subsubsection{Case C: Real part equal to imaginary part ($\beta=\gamma$)}

In this case, the solution diverges; however, as we will show below, $\beta=\gamma$
corresponds to a purely repulsive potential so one could expect such singular behavior. 
Since here $\beta=\gamma$, we can rewrite
\begin{equation}
u=\beta+i\beta=|u|e^{i\pi/4}. \label{u1}
\end{equation}
Also, since $u=(mk/\hbar^2)^{1/4}x$ and $k=|k|e^{i\theta}$, we have
\begin{eqnarray}
u&=&k^{1/4}(m/\hbar^2)^{1/4}x\\
&=&e^{i\theta/4}(m|k|/\hbar^2)^{1/4}x \label{u2}
\end{eqnarray}
Equating the imaginary parts of Equations (\ref{u1}) and (\ref{u2}), we have
\begin{equation}
e^{i\pi/4}=e^{i\theta/4}
\end{equation}
which means $\theta=\pi$ and $k=-1|k|$. This corresponds to a repulsive spring constant 
which has no bound solutions.

\subsection{Comparison of Numerical Solution and Analytic Solution}

In this section, we will examine two different complex values of $k$.  Since $u=\alpha x$ 
where $\alpha=(mk/\hbar^2)^{1/4}$, the solution depends on the real and imaginary parts 
of the spring constant $k$. This determines whether $A=0$ or $B=0$ in Eq. (\ref{gensoln})
as discussed above.
Below, we use natural units with $\hbar=c=1$ and take $m=1$ GeV.

\subsubsection{Case $k = 1 + i$}

In this case, we have $k=e^{i\pi/4}|k|$ and 
$u=e^{i\pi\theta/16}u_r$ where $u_r=(m|k|/\hbar^2)^{1/4}x$. 
Therefore, $\beta>\gamma$ and we must use case A. The solution becomes

\begin{equation}
\psi(u)=A\,U \left( -\frac{\lambda}{2},\sqrt{2} u\right) .
\label{case1sol}
\end{equation}

\begin{figure}[tbp]
\includegraphics[width=16cm]{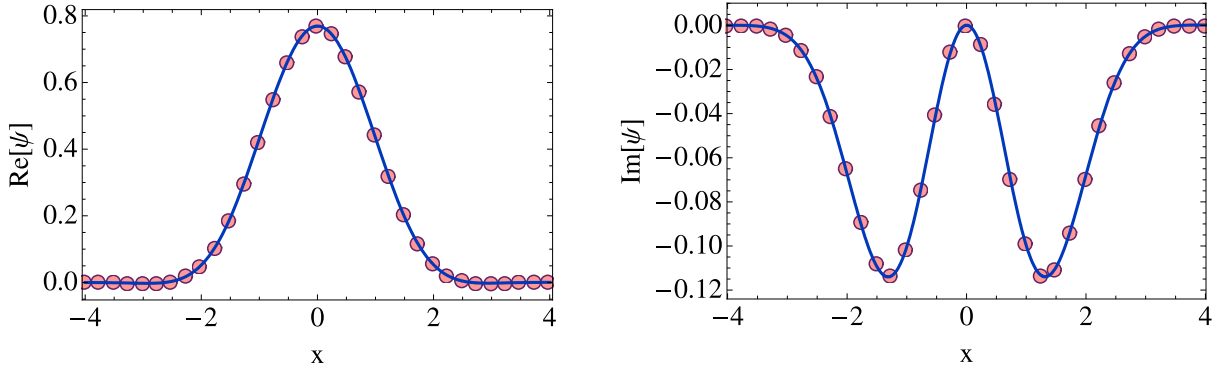}
\caption{Plot showing real and imaginary parts of the wavefunction for a complex harmonic oscillator potential with $k = 1 + i$.  
The solid lines are the analytic results and the circles are sampled points obtained from the numerical solution using
the FDTD method \cite{Strickland:2009ft}.}
\label{fig:app1}
\end{figure}

In Fig.~\ref{fig:app1} we plot the result for the ground state which corresponds to $\lambda=1$ with the
constant $A$
fixed to require a normalized wavefunction.  The
solid lines are the analytic results and the circles are sampled points obtained from numerical solution using
the FDTD method \cite{Strickland:2009ft}.\footnote{The wavefunction can be rotated by an arbitrary
complex phase $e^{i \phi}$ without affecting the probability amplitudes.   In practice, the code converges
to a different random phase angle during each run.   In order to compare to the analytic
results for the real and imaginary parts of the wavefunction 
we manually rotate the numerically determined wavefunctions such that ${\rm Im}[\psi(x=0)]=0$.}
As we can see from this figure, the FDTD algorithm is able
to obtain very good agreement in both the real and imaginary parts of the wavefunction.  We can also
compare the ground state energy predicted analytically by Eq.~(\ref{case1sol}) which is $E_0 = 0.549342+ 0.227545 i$
with the FDTD algorithm's result.
The FDTD algorithm, using 200 points distributed with a step size of $0.05$ and a convergence tolerance
of $10^{-8}$, gives $E_0 = 0.549251+0.227479 i$, which represents an accuracy of approximately 0.01\%.

\subsubsection{Case $k = -i$}

In this case, we have $k=e^{i3\pi/2}|k|$ and 
$u=e^{i3\pi\theta/8}u_r$ where $u_r=(m|k|/\hbar^2)^{1/4}x$. 
Therefore, $\beta<\gamma$ and we must use case B. The solution becomes

\begin{equation}
\psi(u)=B\,U\left(\frac{\lambda}{2}, i\sqrt{2} u \right) .
\label{case2sol}
\end{equation}

\begin{figure}[tbp]
\includegraphics[width=16cm]{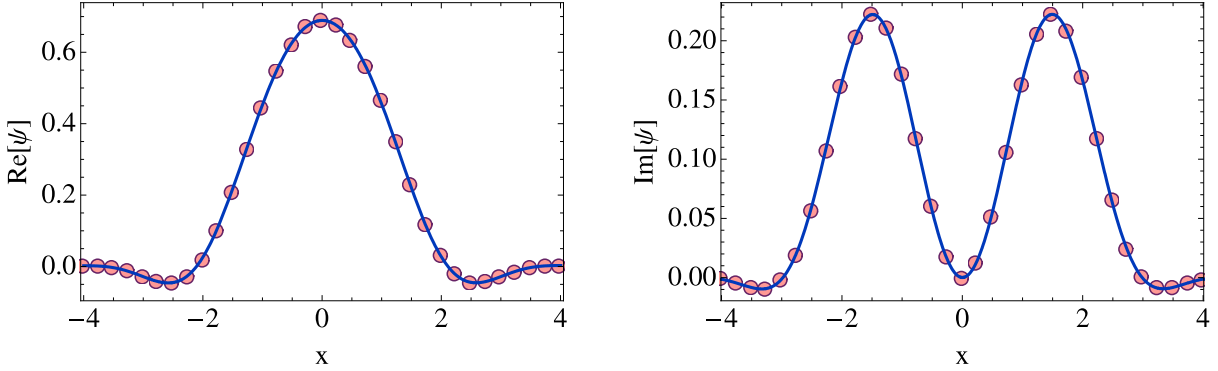}
\caption{Plot showing real and imaginary parts of the wavefunction for a complex harmonic oscillator potential with $k = -i$.  
The solid lines are the analytic results and the circles are sampled points obtained from numerical solution using
the FDTD method \cite{Strickland:2009ft}.}
\label{fig:app2}
\end{figure}

In Fig.~\ref{fig:app2} we plot the result for the ground state which corresponds to $\lambda=1$ with $B$
fixed to require a normalized wavefunction.  The
solid lines are the analytic results and the circles are sampled points obtained from numerical solution using
the FDTD method \cite{Strickland:2009ft}.  
As we can see from this figure, the FDTD algorithm is able
to obtain very good agreement in both the real and imaginary parts of the wavefunction.  We can also
compare the ground state energy predicted analytically by Eq.~(\ref{case2sol}) which is $E_0 = 0.353553- 0.353553 i$
with the FDTD algorithm's result.
The FDTD algorithm, using 200 points distributed with step size of $0.05$ and a convergence tolerance
of $10^{-8}$, gives $E_0 = 0.353522-0.353478 i$, which represents an accuracy of approximately $5\times10^{-3}$ \%.

The results of the two cases presented in this appendix show that the FDTD algorithm
is able to obtain accurate wavefunctions and eigenvalues even in the case that the potential
is complex-valued.  The agreement between the analytic and numerical results can be 
improved by using finer lattice spacings and larger number of points.

\bibliography{complexpotential}

\begin{thebibliography}{65}
\expandafter\ifx\csname natexlab\endcsname\relax\def\natexlab#1{#1}\fi
\expandafter\ifx\csname bibnamefont\endcsname\relax
  \def\bibnamefont#1{#1}\fi
\expandafter\ifx\csname bibfnamefont\endcsname\relax
  \def\bibfnamefont#1{#1}\fi
\expandafter\ifx\csname citenamefont\endcsname\relax
  \def\citenamefont#1{#1}\fi
\expandafter\ifx\csname url\endcsname\relax
  \def\url#1{\texttt{#1}}\fi
\expandafter\ifx\csname urlprefix\endcsname\relax\def\urlprefix{URL }\fi
\providecommand{\bibinfo}[2]{#2}
\providecommand{\eprint}[2][]{\url{#2}}

\bibitem[{\citenamefont{Shuryak}(1980)}]{Shuryak:1980tp}
\bibinfo{author}{\bibfnamefont{E.~V.} \bibnamefont{Shuryak}},
  \bibinfo{journal}{Phys. Rept.} \textbf{\bibinfo{volume}{61}},
  \bibinfo{pages}{71} (\bibinfo{year}{1980}).

\bibitem[{\citenamefont{Gross et~al.}(1981)\citenamefont{Gross, Pisarski, and
  Yaffe}}]{Gross:1980br}
\bibinfo{author}{\bibfnamefont{D.~J.} \bibnamefont{Gross}},
  \bibinfo{author}{\bibfnamefont{R.~D.} \bibnamefont{Pisarski}},
  \bibnamefont{and} \bibinfo{author}{\bibfnamefont{L.~G.} \bibnamefont{Yaffe}},
  \bibinfo{journal}{Rev. Mod. Phys.} \textbf{\bibinfo{volume}{53}},
  \bibinfo{pages}{43} (\bibinfo{year}{1981}).

\bibitem[{\citenamefont{Lucha et~al.}(1991)\citenamefont{Lucha, Schoberl, and
  Gromes}}]{Lucha:1991vn}
\bibinfo{author}{\bibfnamefont{W.}~\bibnamefont{Lucha}},
  \bibinfo{author}{\bibfnamefont{F.~F.} \bibnamefont{Schoberl}},
  \bibnamefont{and} \bibinfo{author}{\bibfnamefont{D.}~\bibnamefont{Gromes}},
  \bibinfo{journal}{Phys. Rept.} \textbf{\bibinfo{volume}{200}},
  \bibinfo{pages}{127} (\bibinfo{year}{1991}).

\bibitem[{\citenamefont{Eichten et~al.}(1980)\citenamefont{Eichten, Gottfried,
  Kinoshita, Lane, and Yan}}]{Eichten:1979ms}
\bibinfo{author}{\bibfnamefont{E.}~\bibnamefont{Eichten}},
  \bibinfo{author}{\bibfnamefont{K.}~\bibnamefont{Gottfried}},
  \bibinfo{author}{\bibfnamefont{T.}~\bibnamefont{Kinoshita}},
  \bibinfo{author}{\bibfnamefont{K.~D.} \bibnamefont{Lane}}, \bibnamefont{and}
  \bibinfo{author}{\bibfnamefont{T.-M.} \bibnamefont{Yan}},
  \bibinfo{journal}{Phys. Rev.} \textbf{\bibinfo{volume}{D21}},
  \bibinfo{pages}{203} (\bibinfo{year}{1980}).

\bibitem[{\citenamefont{Brambilla et~al.}(2005)\citenamefont{Brambilla, Pineda,
  Soto, and Vairo}}]{Brambilla:2004jw}
\bibinfo{author}{\bibfnamefont{N.}~\bibnamefont{Brambilla}},
  \bibinfo{author}{\bibfnamefont{A.}~\bibnamefont{Pineda}},
  \bibinfo{author}{\bibfnamefont{J.}~\bibnamefont{Soto}}, \bibnamefont{and}
  \bibinfo{author}{\bibfnamefont{A.}~\bibnamefont{Vairo}},
  \bibinfo{journal}{Rev. Mod. Phys.} \textbf{\bibinfo{volume}{77}},
  \bibinfo{pages}{1423} (\bibinfo{year}{2005}), \eprint{hep-ph/0410047}.

\bibitem[{\citenamefont{Matsui and Satz}(1986)}]{Matsui:1986dk}
\bibinfo{author}{\bibfnamefont{T.}~\bibnamefont{Matsui}} \bibnamefont{and}
  \bibinfo{author}{\bibfnamefont{H.}~\bibnamefont{Satz}},
  \bibinfo{journal}{Phys. Lett.} \textbf{\bibinfo{volume}{B178}},
  \bibinfo{pages}{416} (\bibinfo{year}{1986}).

\bibitem[{\citenamefont{Karsch et~al.}(1988)\citenamefont{Karsch, Mehr, and
  Satz}}]{Karsch:1987pv}
\bibinfo{author}{\bibfnamefont{F.}~\bibnamefont{Karsch}},
  \bibinfo{author}{\bibfnamefont{M.~T.} \bibnamefont{Mehr}}, \bibnamefont{and}
  \bibinfo{author}{\bibfnamefont{H.}~\bibnamefont{Satz}}, \bibinfo{journal}{Z.
  Phys.} \textbf{\bibinfo{volume}{C37}}, \bibinfo{pages}{617}
  (\bibinfo{year}{1988}).

\bibitem[{\citenamefont{Mocsy and Petreczky}(2005)}]{Mocsy:2004bv}
\bibinfo{author}{\bibfnamefont{A.}~\bibnamefont{Mocsy}} \bibnamefont{and}
  \bibinfo{author}{\bibfnamefont{P.}~\bibnamefont{Petreczky}},
  \bibinfo{journal}{Eur. Phys. J.} \textbf{\bibinfo{volume}{C43}},
  \bibinfo{pages}{77} (\bibinfo{year}{2005}), \eprint{hep-ph/0411262}.

\bibitem[{\citenamefont{Wong}(2005)}]{Wong:2004zr}
\bibinfo{author}{\bibfnamefont{C.-Y.} \bibnamefont{Wong}},
  \bibinfo{journal}{Phys. Rev.} \textbf{\bibinfo{volume}{C72}},
  \bibinfo{pages}{034906} (\bibinfo{year}{2005}), \eprint{hep-ph/0408020}.

\bibitem[{\citenamefont{Mocsy and Petreczky}(2006)}]{Mocsy:2005qw}
\bibinfo{author}{\bibfnamefont{A.}~\bibnamefont{Mocsy}} \bibnamefont{and}
  \bibinfo{author}{\bibfnamefont{P.}~\bibnamefont{Petreczky}},
  \bibinfo{journal}{Phys. Rev.} \textbf{\bibinfo{volume}{D73}},
  \bibinfo{pages}{074007} (\bibinfo{year}{2006}), \eprint{hep-ph/0512156}.

\bibitem[{\citenamefont{Cabrera and Rapp}(2007)}]{Cabrera:2006wh}
\bibinfo{author}{\bibfnamefont{D.}~\bibnamefont{Cabrera}} \bibnamefont{and}
  \bibinfo{author}{\bibfnamefont{R.}~\bibnamefont{Rapp}},
  \bibinfo{journal}{Phys. Rev.} \textbf{\bibinfo{volume}{D76}},
  \bibinfo{pages}{114506} (\bibinfo{year}{2007}), \eprint{hep-ph/0611134}.

\bibitem[{\citenamefont{Mocsy and Petreczky}(2007)}]{Mocsy:2007jz}
\bibinfo{author}{\bibfnamefont{A.}~\bibnamefont{Mocsy}} \bibnamefont{and}
  \bibinfo{author}{\bibfnamefont{P.}~\bibnamefont{Petreczky}},
  \bibinfo{journal}{Phys. Rev. Lett.} \textbf{\bibinfo{volume}{99}},
  \bibinfo{pages}{211602} (\bibinfo{year}{2007}), \eprint{0706.2183}.

\bibitem[{\citenamefont{Alberico et~al.}(2008)\citenamefont{Alberico, Beraudo,
  De~Pace, and Molinari}}]{Alberico:2007rg}
\bibinfo{author}{\bibfnamefont{W.~M.} \bibnamefont{Alberico}},
  \bibinfo{author}{\bibfnamefont{A.}~\bibnamefont{Beraudo}},
  \bibinfo{author}{\bibfnamefont{A.}~\bibnamefont{De~Pace}}, \bibnamefont{and}
  \bibinfo{author}{\bibfnamefont{A.}~\bibnamefont{Molinari}},
  \bibinfo{journal}{Phys. Rev.} \textbf{\bibinfo{volume}{D77}},
  \bibinfo{pages}{017502} (\bibinfo{year}{2008}), \eprint{0706.2846}.

\bibitem[{\citenamefont{Mocsy and Petreczky}(2008)}]{Mocsy:2007yj}
\bibinfo{author}{\bibfnamefont{A.}~\bibnamefont{Mocsy}} \bibnamefont{and}
  \bibinfo{author}{\bibfnamefont{P.}~\bibnamefont{Petreczky}},
  \bibinfo{journal}{Phys. Rev.} \textbf{\bibinfo{volume}{D77}},
  \bibinfo{pages}{014501} (\bibinfo{year}{2008}), \eprint{0705.2559}.

\bibitem[{\citenamefont{Mocsy}(2008)}]{Mocsy:2008eg}
\bibinfo{author}{\bibfnamefont{A.}~\bibnamefont{Mocsy}} (\bibinfo{year}{2008}),
  \eprint{0811.0337}.

\bibitem[{\citenamefont{Umeda et~al.}(2005)\citenamefont{Umeda, Nomura, and
  Matsufuru}}]{Umeda:2002vr}
\bibinfo{author}{\bibfnamefont{T.}~\bibnamefont{Umeda}},
  \bibinfo{author}{\bibfnamefont{K.}~\bibnamefont{Nomura}}, \bibnamefont{and}
  \bibinfo{author}{\bibfnamefont{H.}~\bibnamefont{Matsufuru}},
  \bibinfo{journal}{Eur. Phys. J.} \textbf{\bibinfo{volume}{C39S1}},
  \bibinfo{pages}{9} (\bibinfo{year}{2005}), \eprint{hep-lat/0211003}.

\bibitem[{\citenamefont{Asakawa and Hatsuda}(2004)}]{Asakawa:2003re}
\bibinfo{author}{\bibfnamefont{M.}~\bibnamefont{Asakawa}} \bibnamefont{and}
  \bibinfo{author}{\bibfnamefont{T.}~\bibnamefont{Hatsuda}},
  \bibinfo{journal}{Phys. Rev. Lett.} \textbf{\bibinfo{volume}{92}},
  \bibinfo{pages}{012001} (\bibinfo{year}{2004}), \eprint{hep-lat/0308034}.

\bibitem[{\citenamefont{Datta et~al.}(2004)\citenamefont{Datta, Karsch,
  Petreczky, and Wetzorke}}]{Datta:2003ww}
\bibinfo{author}{\bibfnamefont{S.}~\bibnamefont{Datta}},
  \bibinfo{author}{\bibfnamefont{F.}~\bibnamefont{Karsch}},
  \bibinfo{author}{\bibfnamefont{P.}~\bibnamefont{Petreczky}},
  \bibnamefont{and} \bibinfo{author}{\bibfnamefont{I.}~\bibnamefont{Wetzorke}},
  \bibinfo{journal}{Phys. Rev.} \textbf{\bibinfo{volume}{D69}},
  \bibinfo{pages}{094507} (\bibinfo{year}{2004}), \eprint{hep-lat/0312037}.

\bibitem[{\citenamefont{Aarts et~al.}(2006)}]{Aarts:2006nr}
\bibinfo{author}{\bibfnamefont{G.}~\bibnamefont{Aarts}} \bibnamefont{et~al.},
  \bibinfo{journal}{PoS} \textbf{\bibinfo{volume}{LAT2006}},
  \bibinfo{pages}{126} (\bibinfo{year}{2006}), \eprint{hep-lat/0610065}.

\bibitem[{\citenamefont{Hatsuda}(2006)}]{Hatsuda:2006zz}
\bibinfo{author}{\bibfnamefont{T.}~\bibnamefont{Hatsuda}},
  \bibinfo{journal}{PoS} \textbf{\bibinfo{volume}{LAT2006}},
  \bibinfo{pages}{010} (\bibinfo{year}{2006}).

\bibitem[{\citenamefont{Jakovac et~al.}(2007)\citenamefont{Jakovac, Petreczky,
  Petrov, and Velytsky}}]{Jakovac:2006sf}
\bibinfo{author}{\bibfnamefont{A.}~\bibnamefont{Jakovac}},
  \bibinfo{author}{\bibfnamefont{P.}~\bibnamefont{Petreczky}},
  \bibinfo{author}{\bibfnamefont{K.}~\bibnamefont{Petrov}}, \bibnamefont{and}
  \bibinfo{author}{\bibfnamefont{A.}~\bibnamefont{Velytsky}},
  \bibinfo{journal}{Phys. Rev.} \textbf{\bibinfo{volume}{D75}},
  \bibinfo{pages}{014506} (\bibinfo{year}{2007}), \eprint{hep-lat/0611017}.

\bibitem[{\citenamefont{Umeda}(2007)}]{Umeda:2007hy}
\bibinfo{author}{\bibfnamefont{T.}~\bibnamefont{Umeda}},
  \bibinfo{journal}{Phys. Rev.} \textbf{\bibinfo{volume}{D75}},
  \bibinfo{pages}{094502} (\bibinfo{year}{2007}), \eprint{hep-lat/0701005}.

\bibitem[{\citenamefont{Aarts et~al.}(2007)\citenamefont{Aarts, Allton, Oktay,
  Peardon, and Skullerud}}]{Aarts:2007pk}
\bibinfo{author}{\bibfnamefont{G.}~\bibnamefont{Aarts}},
  \bibinfo{author}{\bibfnamefont{C.}~\bibnamefont{Allton}},
  \bibinfo{author}{\bibfnamefont{M.~B.} \bibnamefont{Oktay}},
  \bibinfo{author}{\bibfnamefont{M.}~\bibnamefont{Peardon}}, \bibnamefont{and}
  \bibinfo{author}{\bibfnamefont{J.-I.} \bibnamefont{Skullerud}},
  \bibinfo{journal}{Phys. Rev.} \textbf{\bibinfo{volume}{D76}},
  \bibinfo{pages}{094513} (\bibinfo{year}{2007}), \eprint{0705.2198}.

\bibitem[{\citenamefont{Aarts et~al.}(2010)}]{Aarts:2010ek}
\bibinfo{author}{\bibfnamefont{G.}~\bibnamefont{Aarts}} \bibnamefont{et~al.}
  (\bibinfo{year}{2010}), \eprint{1010.3725}.

\bibitem[{\citenamefont{Nakahara et~al.}(1999)\citenamefont{Nakahara, Asakawa,
  and Hatsuda}}]{Nakahara:1999vy}
\bibinfo{author}{\bibfnamefont{Y.}~\bibnamefont{Nakahara}},
  \bibinfo{author}{\bibfnamefont{M.}~\bibnamefont{Asakawa}}, \bibnamefont{and}
  \bibinfo{author}{\bibfnamefont{T.}~\bibnamefont{Hatsuda}},
  \bibinfo{journal}{Phys.Rev.} \textbf{\bibinfo{volume}{D60}},
  \bibinfo{pages}{091503} (\bibinfo{year}{1999}), \eprint{hep-lat/9905034}.

\bibitem[{\citenamefont{Asakawa et~al.}(2001)\citenamefont{Asakawa, Hatsuda,
  and Nakahara}}]{Asakawa:2000tr}
\bibinfo{author}{\bibfnamefont{M.}~\bibnamefont{Asakawa}},
  \bibinfo{author}{\bibfnamefont{T.}~\bibnamefont{Hatsuda}}, \bibnamefont{and}
  \bibinfo{author}{\bibfnamefont{Y.}~\bibnamefont{Nakahara}},
  \bibinfo{journal}{Prog.Part.Nucl.Phys.} \textbf{\bibinfo{volume}{46}},
  \bibinfo{pages}{459} (\bibinfo{year}{2001}), \eprint{hep-lat/0011040}.

\bibitem[{\citenamefont{Asakawa et~al.}(2003)\citenamefont{Asakawa, Hatsuda,
  and Nakahara}}]{Asakawa:2002xj}
\bibinfo{author}{\bibfnamefont{M.}~\bibnamefont{Asakawa}},
  \bibinfo{author}{\bibfnamefont{T.}~\bibnamefont{Hatsuda}}, \bibnamefont{and}
  \bibinfo{author}{\bibfnamefont{Y.}~\bibnamefont{Nakahara}},
  \bibinfo{journal}{Nucl.Phys.} \textbf{\bibinfo{volume}{A715}},
  \bibinfo{pages}{863} (\bibinfo{year}{2003}), \eprint{hep-lat/0208059}.

\bibitem[{\citenamefont{Armesto et~al.}(2008)}]{Abreu:2007kv}
\bibinfo{author}{\bibfnamefont{N.}~\bibnamefont{Armesto}} \bibnamefont{et~al.},
  \bibinfo{journal}{J. Phys.} \textbf{\bibinfo{volume}{G35}},
  \bibinfo{pages}{054001} (\bibinfo{year}{2008}), \eprint{0711.0974}.

\bibitem[{\citenamefont{Rapp et~al.}(2008)\citenamefont{Rapp, Blaschke, and
  Crochet}}]{Rapp:2008tf}
\bibinfo{author}{\bibfnamefont{R.}~\bibnamefont{Rapp}},
  \bibinfo{author}{\bibfnamefont{D.}~\bibnamefont{Blaschke}}, \bibnamefont{and}
  \bibinfo{author}{\bibfnamefont{P.}~\bibnamefont{Crochet}}
  (\bibinfo{year}{2008}), \eprint{0807.2470}.

\bibitem[{\citenamefont{Laine et~al.}(2007{\natexlab{a}})\citenamefont{Laine,
  Philipsen, Romatschke, and Tassler}}]{Laine:2006ns}
\bibinfo{author}{\bibfnamefont{M.}~\bibnamefont{Laine}},
  \bibinfo{author}{\bibfnamefont{O.}~\bibnamefont{Philipsen}},
  \bibinfo{author}{\bibfnamefont{P.}~\bibnamefont{Romatschke}},
  \bibnamefont{and} \bibinfo{author}{\bibfnamefont{M.}~\bibnamefont{Tassler}},
  \bibinfo{journal}{JHEP} \textbf{\bibinfo{volume}{03}}, \bibinfo{pages}{054}
  (\bibinfo{year}{2007}{\natexlab{a}}), \eprint{hep-ph/0611300}.

\bibitem[{\citenamefont{Laine}(2007)}]{Laine:2007gj}
\bibinfo{author}{\bibfnamefont{M.}~\bibnamefont{Laine}},
  \bibinfo{journal}{JHEP} \textbf{\bibinfo{volume}{05}}, \bibinfo{pages}{028}
  (\bibinfo{year}{2007}), \eprint{0704.1720}.

\bibitem[{\citenamefont{Beraudo et~al.}(2008)\citenamefont{Beraudo, Blaizot,
  and Ratti}}]{Beraudo:2007ky}
\bibinfo{author}{\bibfnamefont{A.}~\bibnamefont{Beraudo}},
  \bibinfo{author}{\bibfnamefont{J.~P.} \bibnamefont{Blaizot}},
  \bibnamefont{and} \bibinfo{author}{\bibfnamefont{C.}~\bibnamefont{Ratti}},
  \bibinfo{journal}{Nucl. Phys.} \textbf{\bibinfo{volume}{A806}},
  \bibinfo{pages}{312} (\bibinfo{year}{2008}), \eprint{0712.4394}.

\bibitem[{\citenamefont{Brambilla et~al.}(2008)\citenamefont{Brambilla,
  Ghiglieri, Vairo, and Petreczky}}]{Brambilla:2008cx}
\bibinfo{author}{\bibfnamefont{N.}~\bibnamefont{Brambilla}},
  \bibinfo{author}{\bibfnamefont{J.}~\bibnamefont{Ghiglieri}},
  \bibinfo{author}{\bibfnamefont{A.}~\bibnamefont{Vairo}}, \bibnamefont{and}
  \bibinfo{author}{\bibfnamefont{P.}~\bibnamefont{Petreczky}},
  \bibinfo{journal}{Phys. Rev.} \textbf{\bibinfo{volume}{D78}},
  \bibinfo{pages}{014017} (\bibinfo{year}{2008}), \eprint{0804.0993}.

\bibitem[{\citenamefont{Dumitru et~al.}(2008)\citenamefont{Dumitru, Guo, and
  Strickland}}]{Dumitru:2007hy}
\bibinfo{author}{\bibfnamefont{A.}~\bibnamefont{Dumitru}},
  \bibinfo{author}{\bibfnamefont{Y.}~\bibnamefont{Guo}}, \bibnamefont{and}
  \bibinfo{author}{\bibfnamefont{M.}~\bibnamefont{Strickland}},
  \bibinfo{journal}{Phys. Lett.} \textbf{\bibinfo{volume}{B662}},
  \bibinfo{pages}{37} (\bibinfo{year}{2008}), \eprint{0711.4722}.

\bibitem[{\citenamefont{Dumitru
  et~al.}(2009{\natexlab{a}})\citenamefont{Dumitru, Guo, Mocsy, and
  Strickland}}]{Dumitru:2009ni}
\bibinfo{author}{\bibfnamefont{A.}~\bibnamefont{Dumitru}},
  \bibinfo{author}{\bibfnamefont{Y.}~\bibnamefont{Guo}},
  \bibinfo{author}{\bibfnamefont{A.}~\bibnamefont{Mocsy}}, \bibnamefont{and}
  \bibinfo{author}{\bibfnamefont{M.}~\bibnamefont{Strickland}},
  \bibinfo{journal}{Phys. Rev.} \textbf{\bibinfo{volume}{D79}},
  \bibinfo{pages}{054019} (\bibinfo{year}{2009}{\natexlab{a}}),
  \eprint{0901.1998}.

\bibitem[{\citenamefont{Burnier et~al.}(2009)\citenamefont{Burnier, Laine, and
  Vepsalainen}}]{Burnier:2009yu}
\bibinfo{author}{\bibfnamefont{Y.}~\bibnamefont{Burnier}},
  \bibinfo{author}{\bibfnamefont{M.}~\bibnamefont{Laine}}, \bibnamefont{and}
  \bibinfo{author}{\bibfnamefont{M.}~\bibnamefont{Vepsalainen}}
  (\bibinfo{year}{2009}), \eprint{0903.3467}.

\bibitem[{\citenamefont{Dumitru
  et~al.}(2009{\natexlab{b}})\citenamefont{Dumitru, Guo, and
  Strickland}}]{Dumitru:2009fy}
\bibinfo{author}{\bibfnamefont{A.}~\bibnamefont{Dumitru}},
  \bibinfo{author}{\bibfnamefont{Y.}~\bibnamefont{Guo}}, \bibnamefont{and}
  \bibinfo{author}{\bibfnamefont{M.}~\bibnamefont{Strickland}},
  \bibinfo{journal}{Phys. Rev.} \textbf{\bibinfo{volume}{D79}},
  \bibinfo{pages}{114003} (\bibinfo{year}{2009}{\natexlab{b}}),
  \eprint{0903.4703}.

\bibitem[{\citenamefont{Noronha and
  Dumitru}(2009{\natexlab{a}})}]{Noronha:2009ia}
\bibinfo{author}{\bibfnamefont{J.}~\bibnamefont{Noronha}} \bibnamefont{and}
  \bibinfo{author}{\bibfnamefont{A.}~\bibnamefont{Dumitru}},
  \bibinfo{journal}{Phys.Rev.} \textbf{\bibinfo{volume}{D80}},
  \bibinfo{pages}{014007} (\bibinfo{year}{2009}{\natexlab{a}}),
  \eprint{0903.2804}.

\bibitem[{\citenamefont{Philipsen and Tassler}(2009)}]{Philipsen:2009wg}
\bibinfo{author}{\bibfnamefont{O.}~\bibnamefont{Philipsen}} \bibnamefont{and}
  \bibinfo{author}{\bibfnamefont{M.}~\bibnamefont{Tassler}}
  (\bibinfo{year}{2009}), \eprint{0908.1746}.

\bibitem[{\citenamefont{Israel}(1976)}]{Israel:1976tn}
\bibinfo{author}{\bibfnamefont{W.}~\bibnamefont{Israel}},
  \bibinfo{journal}{Ann. Phys.} \textbf{\bibinfo{volume}{100}},
  \bibinfo{pages}{310} (\bibinfo{year}{1976}).

\bibitem[{\citenamefont{Israel and Stewart}(1979)}]{Israel:1979wp}
\bibinfo{author}{\bibfnamefont{W.}~\bibnamefont{Israel}} \bibnamefont{and}
  \bibinfo{author}{\bibfnamefont{J.~M.} \bibnamefont{Stewart}},
  \bibinfo{journal}{Ann. Phys.} \textbf{\bibinfo{volume}{118}},
  \bibinfo{pages}{341} (\bibinfo{year}{1979}).

\bibitem[{\citenamefont{Baym}(1984)}]{Baym:1984np}
\bibinfo{author}{\bibfnamefont{G.}~\bibnamefont{Baym}}, \bibinfo{journal}{Phys.
  Lett.} \textbf{\bibinfo{volume}{B138}}, \bibinfo{pages}{18}
  (\bibinfo{year}{1984}).

\bibitem[{\citenamefont{Muronga}(2002)}]{Muronga:2001zk}
\bibinfo{author}{\bibfnamefont{A.}~\bibnamefont{Muronga}},
  \bibinfo{journal}{Phys. Rev. Lett.} \textbf{\bibinfo{volume}{88}},
  \bibinfo{pages}{062302} (\bibinfo{year}{2002}), \eprint{nucl-th/0104064}.

\bibitem[{\citenamefont{Muronga}(2004)}]{Muronga:2003ta}
\bibinfo{author}{\bibfnamefont{A.}~\bibnamefont{Muronga}},
  \bibinfo{journal}{Phys. Rev.} \textbf{\bibinfo{volume}{C69}},
  \bibinfo{pages}{034903} (\bibinfo{year}{2004}), \eprint{nucl-th/0309055}.

\bibitem[{\citenamefont{Martinez and Strickland}(2009)}]{Martinez:2009mf}
\bibinfo{author}{\bibfnamefont{M.}~\bibnamefont{Martinez}} \bibnamefont{and}
  \bibinfo{author}{\bibfnamefont{M.}~\bibnamefont{Strickland}},
  \bibinfo{journal}{Phys. Rev.} \textbf{\bibinfo{volume}{C79}},
  \bibinfo{pages}{044903} (\bibinfo{year}{2009}), \eprint{0902.3834}.

\bibitem[{\citenamefont{Florkowski and Ryblewski}(2010)}]{Florkowski:2010cf}
\bibinfo{author}{\bibfnamefont{W.}~\bibnamefont{Florkowski}} \bibnamefont{and}
  \bibinfo{author}{\bibfnamefont{R.}~\bibnamefont{Ryblewski}}
  (\bibinfo{year}{2010}), \eprint{1007.0130}.

\bibitem[{\citenamefont{Martinez and
  Strickland}(2010{\natexlab{a}})}]{Martinez:2010sc}
\bibinfo{author}{\bibfnamefont{M.}~\bibnamefont{Martinez}} \bibnamefont{and}
  \bibinfo{author}{\bibfnamefont{M.}~\bibnamefont{Strickland}},
  \bibinfo{journal}{Nucl. Phys.} \textbf{\bibinfo{volume}{A848}},
  \bibinfo{pages}{183} (\bibinfo{year}{2010}{\natexlab{a}}),
  \eprint{1007.0889}.

\bibitem[{\citenamefont{Martinez and
  Strickland}(2010{\natexlab{b}})}]{Martinez:2010sd}
\bibinfo{author}{\bibfnamefont{M.}~\bibnamefont{Martinez}} \bibnamefont{and}
  \bibinfo{author}{\bibfnamefont{M.}~\bibnamefont{Strickland}}
  (\bibinfo{year}{2010}{\natexlab{b}}), \eprint{1011.3056}.

\bibitem[{\citenamefont{Dumitru}(2010)}]{Dumitru:2010id}
\bibinfo{author}{\bibfnamefont{A.}~\bibnamefont{Dumitru}}
  (\bibinfo{year}{2010}), \eprint{1010.5218}.

\bibitem[{\citenamefont{Romatschke and Strickland}(2003)}]{Romatschke:2003ms}
\bibinfo{author}{\bibfnamefont{P.}~\bibnamefont{Romatschke}} \bibnamefont{and}
  \bibinfo{author}{\bibfnamefont{M.}~\bibnamefont{Strickland}},
  \bibinfo{journal}{Phys. Rev.} \textbf{\bibinfo{volume}{D68}},
  \bibinfo{pages}{036004} (\bibinfo{year}{2003}), \eprint{hep-ph/0304092}.

\bibitem[{\citenamefont{Mrowczynski et~al.}(2004)\citenamefont{Mrowczynski,
  Rebhan, and Strickland}}]{Mrowczynski:2004kv}
\bibinfo{author}{\bibfnamefont{S.}~\bibnamefont{Mrowczynski}},
  \bibinfo{author}{\bibfnamefont{A.}~\bibnamefont{Rebhan}}, \bibnamefont{and}
  \bibinfo{author}{\bibfnamefont{M.}~\bibnamefont{Strickland}},
  \bibinfo{journal}{Phys. Rev.} \textbf{\bibinfo{volume}{D70}},
  \bibinfo{pages}{025004} (\bibinfo{year}{2004}), \eprint{hep-ph/0403256}.

\bibitem[{\citenamefont{Romatschke and Strickland}(2004)}]{Romatschke:2004jh}
\bibinfo{author}{\bibfnamefont{P.}~\bibnamefont{Romatschke}} \bibnamefont{and}
  \bibinfo{author}{\bibfnamefont{M.}~\bibnamefont{Strickland}},
  \bibinfo{journal}{Phys. Rev.} \textbf{\bibinfo{volume}{D70}},
  \bibinfo{pages}{116006} (\bibinfo{year}{2004}), \eprint{hep-ph/0406188}.

\bibitem[{\citenamefont{Schenke and Strickland}(2006)}]{Schenke:2006fz}
\bibinfo{author}{\bibfnamefont{B.}~\bibnamefont{Schenke}} \bibnamefont{and}
  \bibinfo{author}{\bibfnamefont{M.}~\bibnamefont{Strickland}},
  \bibinfo{journal}{Phys. Rev.} \textbf{\bibinfo{volume}{D74}},
  \bibinfo{pages}{065004} (\bibinfo{year}{2006}), \eprint{hep-ph/0606160}.

\bibitem[{\citenamefont{Asakawa et~al.}(2007)\citenamefont{Asakawa, Bass, and
  Muller}}]{Asakawa:2006jn}
\bibinfo{author}{\bibfnamefont{M.}~\bibnamefont{Asakawa}},
  \bibinfo{author}{\bibfnamefont{S.~A.} \bibnamefont{Bass}}, \bibnamefont{and}
  \bibinfo{author}{\bibfnamefont{B.}~\bibnamefont{Muller}},
  \bibinfo{journal}{Prog. Theor. Phys.} \textbf{\bibinfo{volume}{116}},
  \bibinfo{pages}{725} (\bibinfo{year}{2007}), \eprint{hep-ph/0608270}.

\bibitem[{\citenamefont{Laine et~al.}(2007{\natexlab{b}})\citenamefont{Laine,
  Philipsen, and Tassler}}]{Laine:2007qy}
\bibinfo{author}{\bibfnamefont{M.}~\bibnamefont{Laine}},
  \bibinfo{author}{\bibfnamefont{O.}~\bibnamefont{Philipsen}},
  \bibnamefont{and} \bibinfo{author}{\bibfnamefont{M.}~\bibnamefont{Tassler}},
  \bibinfo{journal}{JHEP} \textbf{\bibinfo{volume}{09}}, \bibinfo{pages}{066}
  (\bibinfo{year}{2007}{\natexlab{b}}), \eprint{0707.2458}.

\bibitem[{\citenamefont{Kaczmarek et~al.}(2004)\citenamefont{Kaczmarek, Karsch,
  Zantow, and Petreczky}}]{Kaczmarek:2004gv}
\bibinfo{author}{\bibfnamefont{O.}~\bibnamefont{Kaczmarek}},
  \bibinfo{author}{\bibfnamefont{F.}~\bibnamefont{Karsch}},
  \bibinfo{author}{\bibfnamefont{F.}~\bibnamefont{Zantow}}, \bibnamefont{and}
  \bibinfo{author}{\bibfnamefont{P.}~\bibnamefont{Petreczky}},
  \bibinfo{journal}{Phys. Rev.} \textbf{\bibinfo{volume}{D70}},
  \bibinfo{pages}{074505} (\bibinfo{year}{2004}), \eprint{hep-lat/0406036}.

\bibitem[{\citenamefont{Andersen
  et~al.}(2011{\natexlab{a}})\citenamefont{Andersen, Leganger, Strickland, and
  Su}}]{Andersen:2010wu}
\bibinfo{author}{\bibfnamefont{J.~O.} \bibnamefont{Andersen}},
  \bibinfo{author}{\bibfnamefont{L.~E.} \bibnamefont{Leganger}},
  \bibinfo{author}{\bibfnamefont{M.}~\bibnamefont{Strickland}},
  \bibnamefont{and} \bibinfo{author}{\bibfnamefont{N.}~\bibnamefont{Su}},
  \bibinfo{journal}{Phys. Lett.} \textbf{\bibinfo{volume}{B696}},
  \bibinfo{pages}{468} (\bibinfo{year}{2011}{\natexlab{a}}),
  \eprint{1009.4644}.

\bibitem[{\citenamefont{Andersen
  et~al.}(2011{\natexlab{b}})\citenamefont{Andersen, Leganger, Strickland, and
  Su}}]{Andersen:2011sf}
\bibinfo{author}{\bibfnamefont{J.~O.} \bibnamefont{Andersen}},
  \bibinfo{author}{\bibfnamefont{L.~E.} \bibnamefont{Leganger}},
  \bibinfo{author}{\bibfnamefont{M.}~\bibnamefont{Strickland}},
  \bibnamefont{and} \bibinfo{author}{\bibfnamefont{N.}~\bibnamefont{Su}}
  (\bibinfo{year}{2011}{\natexlab{b}}), \eprint{1103.2528}.

\bibitem[{\citenamefont{Andersen
  et~al.}(2010{\natexlab{a}})\citenamefont{Andersen, Strickland, and
  Su}}]{Andersen:2010ct}
\bibinfo{author}{\bibfnamefont{J.~O.} \bibnamefont{Andersen}},
  \bibinfo{author}{\bibfnamefont{M.}~\bibnamefont{Strickland}},
  \bibnamefont{and} \bibinfo{author}{\bibfnamefont{N.}~\bibnamefont{Su}},
  \bibinfo{journal}{JHEP} \textbf{\bibinfo{volume}{08}}, \bibinfo{pages}{113}
  (\bibinfo{year}{2010}{\natexlab{a}}), \eprint{1005.1603}.

\bibitem[{\citenamefont{Andersen
  et~al.}(2010{\natexlab{b}})\citenamefont{Andersen, Strickland, and
  Su}}]{Andersen:2009tc}
\bibinfo{author}{\bibfnamefont{J.~O.} \bibnamefont{Andersen}},
  \bibinfo{author}{\bibfnamefont{M.}~\bibnamefont{Strickland}},
  \bibnamefont{and} \bibinfo{author}{\bibfnamefont{N.}~\bibnamefont{Su}},
  \bibinfo{journal}{Phys. Rev. Lett.} \textbf{\bibinfo{volume}{104}},
  \bibinfo{pages}{122003} (\bibinfo{year}{2010}{\natexlab{b}}),
  \eprint{0911.0676}.

\bibitem[{\citenamefont{Noronha and
  Dumitru}(2009{\natexlab{b}})}]{Noronha:2009da}
\bibinfo{author}{\bibfnamefont{J.}~\bibnamefont{Noronha}} \bibnamefont{and}
  \bibinfo{author}{\bibfnamefont{A.}~\bibnamefont{Dumitru}},
  \bibinfo{journal}{Phys.Rev.Lett.} \textbf{\bibinfo{volume}{103}},
  \bibinfo{pages}{152304} (\bibinfo{year}{2009}{\natexlab{b}}),
  \eprint{0907.3062}.

\bibitem[{\citenamefont{Sudiarta and Geldart}(2007)}]{Sudiarta:2007}
\bibinfo{author}{\bibfnamefont{I.}~\bibnamefont{Sudiarta}} \bibnamefont{and}
  \bibinfo{author}{\bibfnamefont{D.}~\bibnamefont{Geldart}},
  \bibinfo{journal}{Journal of Physics} \textbf{\bibinfo{volume}{A40}},
  \bibinfo{pages}{1885} (\bibinfo{year}{2007}).

\bibitem[{\citenamefont{Strickland and
  Yager-Elorriaga}(2010)}]{Strickland:2009ft}
\bibinfo{author}{\bibfnamefont{M.}~\bibnamefont{Strickland}} \bibnamefont{and}
  \bibinfo{author}{\bibfnamefont{D.}~\bibnamefont{Yager-Elorriaga}},
  \bibinfo{journal}{J. Comput. Phys.} \textbf{\bibinfo{volume}{229}},
  \bibinfo{pages}{6015} (\bibinfo{year}{2010}), \eprint{0904.0939}.

\bibitem[{\citenamefont{Martinez and Strickland}(2008)}]{Martinez:2008di}
\bibinfo{author}{\bibfnamefont{M.}~\bibnamefont{Martinez}} \bibnamefont{and}
  \bibinfo{author}{\bibfnamefont{M.}~\bibnamefont{Strickland}},
  \bibinfo{journal}{Phys. Rev.} \textbf{\bibinfo{volume}{C78}},
  \bibinfo{pages}{034917} (\bibinfo{year}{2008}), \eprint{0805.4552}.

\bibitem[{\citenamefont{{Abramowitz} and {Stegun}}(1964)}]{abramowitz+stegun}
\bibinfo{author}{\bibfnamefont{M.}~\bibnamefont{{Abramowitz}}}
  \bibnamefont{and} \bibinfo{author}{\bibfnamefont{I.~A.}
  \bibnamefont{{Stegun}}}, \emph{\bibinfo{title}{Handbook of Mathematical
  Functions with Formulas, Graphs, and Mathematical Tables}}
  (\bibinfo{publisher}{Dover}, \bibinfo{address}{New York},
  \bibinfo{year}{1964}), \bibinfo{edition}{ninth dover printing, tenth gpo
  printing} ed.

\end{thebibliography}

\end{document}